\documentclass[aps,prx,twocolumn,amsmath,amssymb,superscriptaddress]{revtex4-1}

\usepackage{graphics}
\usepackage{dcolumn}% Align table columns on decimal point
\usepackage{bm}% bold math
\usepackage[caption=false]{subfig} 
\usepackage{color}
\usepackage{tikz}
\usepackage{subfig}
\usepackage[titles,subfigure]{tocloft}
\usepackage{soul}

\begin{document}

\title{Competing states for the fractional quantum Hall effect in the 1/3-filled second Landau level}

\author{Jae-Seung Jeong}
\affiliation{Center for Correlated Electron Systems, Institute for Basic Science, Seoul 08826, Korea}
\affiliation{Department of Physics and Astronomy, Seoul National University, Seoul 08826, Korea}
\author{Hantao Lu}
\affiliation{Center for Interdisciplinary Studies and Key Laboratory for Magnetism and Magnetic Materials of the Ministry of Education, Lanzhou University, Lanzhou 730000, China}
\author{Ki Hoon Lee}
\affiliation{Center for Correlated Electron Systems, Institute for Basic Science, Seoul 08826, Korea}
\affiliation{Department of Physics and Astronomy, Seoul National University, Seoul 08826, Korea}
\author{Kenji Hashimoto}
\affiliation{Max Planck Institute for Mathematics, Vivatsgasse 7, 53111 Bonn, Germany}
\author{Suk Bum Chung}
\affiliation{Center for Correlated Electron Systems, Institute for Basic Science, Seoul 08826, Korea}
\affiliation{Department of Physics and Astronomy, Seoul National University, Seoul 08826, Korea}
\author{Kwon Park}
\email{kpark@kias.re.kr}
\affiliation{School of Physics, Korea Institute for Advanced Study, Seoul 02455, Korea}

\date{\today}
\begin{abstract}
In this work, we investigate the nature of the fractional quantum Hall state in the 1/3-filled second Landau level (SLL) at filling factor $\nu=7/3$ (and 8/3 in the presence of the particle-hole symmetry) via exact diagonalization in both torus and spherical geometries.  
Specifically, we compute the overlap between the exact 7/3 ground state and various competing states including (i) the Laughlin state, (ii) the fermionic Haffnian state, (iii) the antisymmetrized product state of two composite fermion seas at 1/6 filling, and (iv) the particle-hole (PH) conjugate of the $Z_4$ parafermion state. 
All these trial states are constructed according to a guiding principle called the bilayer mapping approach, where a trial state is obtained as the antisymmetrized projection of a bilayer quantum Hall state with interlayer distance $d$ as a variational parameter. 
Under the proper understanding of the ground-state degeneracy in the torus geometry, the $Z_4$ parafermion state can be obtained as the antisymmetrized projection of the Halperin (330) state. 
Similarly, it is proved in this work that the fermionic Haffnian state can be obtained as the antisymmetrized projection of the Halperin (551) state.
The exact 7/3 ground state is obtained as a function of $\delta V_1^{(1)}$, the variation of the first-moment Haldane pseudopotential $V_1^{(1)}$ in the SLL with respect to the pure Coulomb interaction.
It is shown that, while extremely accurate at sufficiently large positive $\delta V_1^{(1)}$, the Laughlin state loses its overlap with the exact 7/3 ground state significantly at $\delta V_1^{(1)} \simeq 0$.
At slightly negative  $\delta V_1^{(1)}$, it is shown that the PH-conjugated $Z_4$ parafermion state has a substantial overlap with the exact 7/3 ground state, which is the highest among the above four trial states. 
Around the Coulomb point, the energy spectrum exhibits an intriguing change from the spectrum with the Laughlin-type magnetoroton structure to that with the specific quasi-degeneracy of the ground state, which is characteristic to the PH-conjugated $Z_4$ parafermion state. 
\end{abstract}

\maketitle

%%%%%%%%%%%%
\section{Introduction}
\label{sec:Introduction}
%%%%%%%%%%%%

The fractional quantum Hall (FQH) states occurring in the second Landau level (SLL) have attracted intense interest due to their possibility as exotic topological states with non-Abelian quasiparticle statistics.
This possibility is in stark contrast to the fact that the major FQH states in the lowest Landau level (LLL) at filling factor $\nu=n/(2pn \pm 1)$ with $p$ and $n$ being positive integers [and their particle-hole (PH) conjugates] can be understood as the weakly-interacting integer quantum Hall states of composite fermion (CF) at effective filling factor $\nu^*=n$, where quasiparticles satisfy Abelian statistics~\cite{Jain89,Jain_Book}. 
The weakly interacting CF theory serves as an excellent guiding principle for the FQH states in the LLL.

Other minor unconventional FQH states in the LLL, for example, occurring at $\nu=4/11$ and $5/13$ can be understood within the extended framework of the CF theory, where CFs form their own FQH states with mixed ``vorticity flavor'' with some carrying two vortices and the other four~\cite{Park00}.
The FQH state at $\nu=3/8$~\cite{Pan03,Pan15} is highly peculiar, but actually related with those occurring at even-denominator filling factors in the SLL~\cite{Mukherjee12}, which requires a new guiding principle as explained below.

The situation is rather complicated in the SLL, where the FQH states are relatively rare and fragile in comparison with the LLL~\cite{Pan08}. 
On the surface, the weakly interacting CF theory seems to work very well.
All odd-denominator FQH fractions in the SLL are well captured by the usual CF sequence $\nu=2+n/(2pn \pm 1)$ and its PH conjugates $\nu=3 \pm n/(2pn \pm 1)$ except for a few, but robust even-denominator FQH states occurring in the half-filled SLL at $\nu=5/2$ $(=2+1/2)$ and $7/2$ $(=3+1/2)$~\cite{Willett87,Eisenstein02,Xia04}, which can be understood as the paired states of CFs~\cite{Scarola00}.
Note that the 7/2 state is the PH conjugate of the 5/2 state in the limit of zero Landau level mixing, in which case the same physics governs both states. 
%For this reason, whenever we mention the 5/2 state from now on, it is implicitly assumed that the same physics is applied to the 7/2 state.  
%The same is true between the 7/3 $(=2+1/3)$ and 8/3 $(=3-1/3)$ states, also between the 12/5 $(=2+2/5)$ and 13/5 $(=3-2/5)$ states, and so on. 

The 5/2 (7/2) state has attracted much attention for the possibility that it may host non-Abelian statistics for low-energy quasiparticles. 
This possibility is largely based on the observation that the exact 5/2 ground state obtained in finite-size numerical studies seems to be well described by the Moore-Read (MR) Pfaffian state~\cite{Moore91,Greiter91,Greiter92} or its PH conjugate known as the anti-Pfaffian state~\cite{Levin07,Lee07} in certain ranges of model parameters.  
Previous finite-size numerical studies utilized various numerical techniques such as exact diagonalization (ED) in both spherical~\cite{Morf98,Peterson08,Moller08,Peterson08_Orbital,Peterson08_Spontaneous,Lu10,Wojs10,Rezayi11,Pakrouski15} and torus geometries~\cite{ Rezayi00,Peterson08_Orbital, Wang09, Peterson08,Papic12,Pakrouski15, Jeong15}, and the density matrix renormalization group (DMRG) method in both spherical~\cite{Feiguin08} and cylindrical~\cite{Zaletel15} geometries.

The true nature of the 5/2 state, however, still remains elusive in part due to the fact that the MR Pfaffian/anti-Pfaffian state breaks the PH symmetry, while the Coulomb interaction preserves it~\cite{Rezayi00,Peterson08_Spontaneous,Wang09}.
The PH symmetry can be broken in real experiments either spontaneously via modified Coulomb interaction~\cite{Peterson08, Peterson08_Spontaneous} or externally via Landau level mixing~\cite{Rezayi11,Simon13,Papic12,Peterson13,Sodemann13,Bishara09,Wojs10,Zaletel15,Pakrouski15}.
Despite this apparent possibility of the PH symmetry breaking in real experiments, it is still problematic that the MR Pfaffian/anti-Pfaffian state has a rather low overlap with the exact 5/2 ground state at the pure Coulomb interaction. 
In fact, it is interesting to mention that, according to a recent study~\cite{Pakrouski15}, the overlap between the MR Pfaffian/anti-Pfaffian and the exact 5/2 ground states is reduced with an increase of Landau level mixing strength, indicating that the Landau level mixing is not necessarily good for the MR Pfaffian/anti-Pfaffian state.

Whether any given trial state is really relevant to the 5/2 FQH physics depends crucially on its overlap with the exact 5/2 ground state.
If there is a trial state, whose overlap with the exact 5/2 ground state is better than that of the MR Pfaffian/anti-Pfaffian state, it could suggest an altogether different mechanism for the 5/2 FQH physics. 
Recently, based on ED in the torus geometry, it was proposed by two of the current authors~\cite{Jeong15} that the exact 5/2 ground state at the pure Coulomb interaction can be better described by the antisymmetrized product state of two CF seas at quarter filling.
This antisymmetrized product state has an additional advantage over the MR Pfaffian/anti-Pfaffian state in that it is susceptible to an anisotropic instability, which is consistent with recent experimental observations~\cite{Liu13,Samkharadze16}.
This state was constructed according to a guiding principle called the bilayer mapping approach, which is explained in detail below.

Similar to the $5/2$ state, the FQH states at $\nu=12/5$ $(=2+2/5)$ and $13/5$ $(=3-2/5)$~\cite{Xia04,Kumar10,Zhang12} have attracted much attention in the context of non-Abelian statistics.
There are several proposed trial states;
(i) the (PH-conjugated) $Z_3$ parafermion state~\cite{Read99}, 
(ii) the Bonderson-Slingerland state~\cite{Bonderson08,Bonderson12}, 
(iii) the hierarchy state~\cite{Haldane83,Halperin84},
(iv) the weakly-interacting CF state~\cite{Jain_Book}, and 
(v) the multipartite CF states~\cite{Sreejith11,Sreejith13}.
Recent numerical studies using the DMRG method in both spherical~\cite{Zhu15} and cylindrical~\cite{Mong15} geometries as well as ED in the torus geometry~\cite{Mong15} suggest that the $13/5$ ($12/5$) state may be in the same universality class as the (PH conjugate of) $Z_3$ parafermion state, which hosts non-Abelian statistics.

In contrast to the 5/2 (7/2) and 12/5 (13/5) states, the FQH state in the 1/3-filled SLL at $\nu=7/3$ (8/3) has attracted relatively little attention. 
One of the reasons for such a negligence is that the 7/3 ground state was generally believed to be in the same universality class as the Laughlin state occurring in the LLL~\cite{DasSarma_Book}.
When scrutinized, however, numerical evidence has not been so conclusive.  
According to an early study using ED in the torus geometry with the hexagonal unit cell~\cite{Haldane_Book}, the exact 7/3 ground state at $N=6$ seemed to be compressible at the pure Coulomb interaction and undergoes a first-order transition to the Laughlin state as the hard-core component of the Haldane pseudopotential increases. 
Moreover, overlap studies in the spherical geometry~\cite{Ambrumenil88,Balram13} have shown that the square of overlap at the pure Coulomb interaction is very low, typically being below $40 \%$ in finite-size systems with $N \leq 12$.

On the other hand, the entanglement spectrum~\cite{Li08} obtained via ED in the spherical geometry~\cite{Balram13,Johri14} and the DMRG method in the infinite cylindrical geometry~\cite{Zaletel15} provides evidence supporting that the $7/3$ state has the Laughlin-type edge excitation spectrum. 
It was argued~\cite{Balram13,Johri14} that the apparent discrepancy between the low ground-state overlap and the Laughlin-type entanglement spectrum could be caused by the fact that quasiparticles/holes in the 7/3 state are very large. 
Large quasiparticles/holes can be well captured by the variational Monte Carlo simulation~\cite{Balram13} as well as the DMRG method~\cite{Johri14}, while not by ED using relatively small finite-size systems.

Despite this argument, however, the substantially low overlap between the exact 7/3 ground and the Laughlin states is alarming and demands a search for a better trial state. 
As mentioned, this situation is rather similar to that between the exact 5/2 ground and the MR Pfaffian/anti-Pfaffian states.   
In this context, an important question is what guiding principle should be used for the FQH states in the SLL.

Considering that the MR Pfaffian/anti-Pfaffian state is generated by the pairing mechanism involving composite fermions, it is plausible that a good trial state for the exact 7/3 ground state can be also generated by a similar ``pairing'' mechanism. %as that of the 5/2 state.
As a generalization of the pairing mechanism responsible for the MR Pfaffian state, Read and Rezayi~\cite{Read99} proposed a guiding principle for the FQH states in higher Landau levels, according to which the FQH states at $\nu=k/(k+2)$ can be generated  as the zero-energy ground state of the fermionic $(k+1)$-body $\delta$-function interaction Hamiltonian, where $k$ is a positive integer. 
The $k=2$ case corresponds to the MR Pfaffian state.
In general, the ground state obtained at a given $k$ is called the $Z_k$ parafermion state, which includes the previously-mentioned $Z_3$ parafermion state at $\nu=12/5$ (13/5).
Physically speaking, the $Z_k$ parafermion state involves $k$-particle clusters, generalizing pairs in the MR Pfaffian state.

Under this guiding principle, the $Z_4$ parafermion state can serve as a natural trial state at $\nu=8/3$ $(=2+2/3)$.  
In fact, a recent numerical work based on ED in the spherical geometry~\cite{Peterson15} has shown that the $Z_4$ parafermion state has a significant overlap with the exact ground state at $\nu=8/3$ in the limit of zero Landau level mixing, where the PH symmetry is preserved. 
In the presence of the PH symmetry, a related trial state can be obtained at $\nu=7/3$ by applying the PH conjugation operator onto the $Z_4$ parafermion state.
%This suggests that, in the presence of the PH symmetry, exactly the same overlap is obtained between the 7/3 state and the PH conjugate of the $Z_4$ parafermion state.
Note that, throughout this work, we do not consider any PH breaking mechanism, and therefore basically the same physical result holds for both $\nu=7/3$ and 8/3, which are related via the PH conjugation.

As an alternative to the $Z_k$ parafermion approach, in this work, we would like to propose a different guiding principle called the bilayer mapping approach~\cite{Jeong15}.
According to this approach, a trial state is constructed as the antisymmetrized projection of a bilayer quantum Hall state with interlayer distance $d$ as a variational parameter.
As mentioned above, this approach had been already applied to the 5/2 FQH problem. 
The MR Pfaffian state is obtained as the antisymmetrized projection of the Halperin (331) state~\cite{Greiter92_Paired}, which occurs at $d/l_B \simeq 1$. 
Another trial state is the antisymmetrized product state of two CF seas at quarter filling, which occurs at $d/l_B \rightarrow \infty$.
The usual CF sea state at half filling is obtained at $d/l_B \rightarrow 0$.
It was found that the antisymmetrized product state of two CF seas at quarter filling has a substantially higher overlap with the exact 5/2 ground state than the MR Pfaffian state at the Coulomb point~\cite{Jeong15}.
This leads to an intriguing question if the bilayer mapping approach can be also applied to the 7/3 FQH problem.

\begin{figure}[t]
\includegraphics[width=0.45\textwidth]{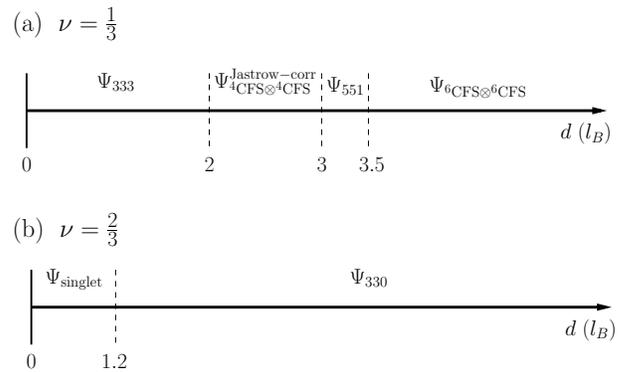}
\caption{
Schematic phase diagrams of the bilayer quantum Hall ground state as a function of interlayer distance $d/l_B$ at (a) $\nu=1/3$ and (b) $2/3$. 
See Sec.~\ref{sec:Trial_states} for the detailed mathematical form of each bilayer quantum Hall ground state.}
\label{fig:bilayer_ground_state}
\end{figure}

To see what trial states can be generated at $\nu=7/3$ in the bilayer mapping approach, let us examine what bilayer quantum Hall ground states can occur as a function of $d/l_B$.
Figure~\ref{fig:bilayer_ground_state} shows the schematic phase diagrams of the bilayer quantum Hall ground state as a function of $d/l_B$ at $\nu=1/3$~\cite{Scarola01} and $2/3$~\cite{McDonald96}.
Scarola and Jain~\cite{Scarola01} studied the phase diagram of the bilayer quantum Hall ground state at $\nu=1/3$ by computing the energies of various trial states as a function of $d/l_B$.
As a result, it was shown that
(i) the Laughlin state, $\Psi_{333}$, has the lowest energy at $0 \leq d/l_B \lesssim 2$, 
(ii) the Jastrow-correlated product state of two CF seas at quarter filling, $\Psi^{\rm Jastrow-corr}_{^4{\rm CFS} \otimes ^4{\rm CFS}}$, at $2 \lesssim d/l_B \lesssim 3$, 
%(ii) the Jastrow correlated product state of two CF seas at quarter filling, $\Psi_{^4{\rm CFS}}\otimes \Psi_{^4{\rm CFS}}$, at $2 \lesssim d/l_B \lesssim 3$, 
(iii) the Halperin (551) state, $\Psi_{551}$, at $3 \lesssim d/l_B \lesssim 3.5$, and finally 
%(iv) the product state of two CF seas at 1/6 filling, $\Psi_{^6{\rm CFS}}\otimes\Psi_{^6{\rm CFS}}$, at $d/l_B \gg 1$.
(iv) the product state of two CF seas at 1/6 filling, $\Psi_{{^6{\rm CFS}}\otimes{^6{\rm CFS}}}$, at $d/l_B \gg 1$.
Meanwhile, McDonald and Haldane~\cite{McDonald96} performed ED to determine the phase diagram of the bilayer quantum Hall ground state at $\nu=2/3$. 
As a result, it was shown that 
(i) the pseudospin singlet state, $\Psi_{\rm singlet}$, occurs at $d/l_B \ll 1$, and 
(ii) the Halperin (330) state, $\Psi_{330}$, at $d/l_B \gg 1$.

Out of these six bilayer quantum Hall ground states, we focus on three states at $\nu=1/3$, which are $\Psi_{333}$, $\Psi_{551}$, and $\Psi_{^6{\rm CFS}\otimes ^6{\rm CFS}}$, and one state at $\nu=2/3$, which is $\Psi_{330}$.
Note that we do not pay attention to $\Psi_{\rm singlet}$ since it is completely annihilated upon antisymmetrization.
Also, we do not discuss $\Psi^{\rm Jastrow-corr}_{^4{\rm CFS} \otimes ^4{\rm CFS}}$ since it turns out that the antisymmetrized projection of this state has a negligible overlap with the exact 7/3 ground state.
We construct the final four trial states for the FQH state at $\nu=7/3$ by applying the antisymmetrization operator onto $\Psi_{333}$, $\Psi_{551}$, $\Psi_{^6{\rm CFS}\otimes ^6{\rm CFS}}$, and by applying the antisymmetrization operator and then the PH conjugation operator onto $\Psi_{330}$.
In summary, we obtain the following four trial states in the bilayer mapping approach:
(i) $\Psi_{333}$, (ii)  ${\cal A} \Psi_{551}$, (iii) ${\cal A} \Psi_{^6{\rm CFS}\otimes ^6{\rm CFS}}$, and (iv) ${\cal C}_{\rm PH} {\cal A} \Psi_{330}$, where ${\cal A}$ and ${\cal C}_{\rm PH}$ are the antisymmetrization and the PH conjugation operators, respectively.
Here, note that ${\cal A} \Psi_{333} = \Psi_{333}$ since the Laughlin state is already antisymmetrized.

%%%%%%%%%%%%%%%%%%%%%%%%%%%%%%%%%%%%%%%%%%%%%%%%%%%%%%%%%%%%%%%%%
%%%%%%%%%%%%%%%%%%%%%%%%%%%%%%%%%%%%%%%%%%%%%%%%%%%%%%%%%%%%%%%%%

While seemingly unrelated, there is in fact an intriguing connection between the two guiding principles of the $Z_k$ parafermion and the bilayer mapping approaches. 
It was shown previously~\cite{Rezayi10,Barkeshli10} that the $Z_4$ parafermion state is entirely equivalent to the antisymmetrized projection of the Halperin (330) state:
\begin{align}
\label{eq:Z4_vs_A330}
\Psi_{Z_4} = {\cal A} \Psi_{330} .
\end{align}
Originally, this identity was derived in an attempt to generate non-Abelian states in bilayer quantum Hall systems as an alternative to the $Z_k$ parafermion approach, which uses the model Hamiltonian of the fermionic $(k+1)$-body $\delta$-function interaction.   
In this point of view, ${\cal A} \Psi_{330}$ can be regarded as the single-layer limit of the bilayer quantum ground state $\Psi_{330}$, which may be obtained in the limit of large interlayer tunneling, while the interlayer Coulomb interaction is set equal to zero. 
Unfortunately, it was shown by a numerical calculation that taking the limit of large interlayer tunneling could be actually different from applying the antisymmetrized projection~\cite{Papic10}.

Similarly, it turns out that there is also an intriguing connection between ${\cal A} \Psi_{551}$ and the previously-known trial state called the fermionic Haffnian state~\cite{Wen94},
%\begin{align}
$\Psi_{\rm Hf} = \Psi_{333} {\rm Det}{\left(\frac{1}{z_i - z_j}\right)}$,
%\end{align} 
which can be regarded as the $d$-wave paired state of composite fermions. 
It is proved in this work that the fermionic Haffnian state is entirely equivalent to the antisymmetrized projection of the Halperin (551) state:
\begin{align}
\label{eq:Hf_vs_A551}
\Psi_{\rm Hf} = {\cal A} \Psi_{551} .
\end{align}
Note that the antisymmetrized projection of the Halperin (551) state has been considered previously~\cite{Regnault08}, although it was not formally identified as the fermionic Haffnian state.
It is interesting to note that there is also a similar connection between the MR Pfaffian state and the antisymmetrized projection of the Halperin (331) state~\cite{Greiter92_Paired}:
\begin{align}
\label{eq:Pf_vs_A331}
\Psi_{\rm Pf} = {\cal A} \Psi_{331} .
\end{align}
Physically, the MR Pfaffian state can be regarded as the $p$-wave paired state of composite fermions.

Proven analytically, all the above identities in Eqs.~\eqref{eq:Z4_vs_A330}, \eqref{eq:Hf_vs_A551}, and \eqref{eq:Pf_vs_A331}, which we call collectively the bilayer mapping identities in this paper, are absolutely exact. 
However, there is a subtlety in applying these identities directly to the torus geometry.
Specifically, the two states equated by the above identities do not necessarily have the same ground-state degeneracy in the torus geometry.
This is in part due to the fact that the bilayer quantum Hall ground states have generally different degeneracies from their counterparts related by the bilayer mapping identities.  
Moreover, the antisymmetrization operator does not necessarily preserve the ground-state degeneracy structure of the bilayer quantum Hall Hamiltonian in the torus geometry.
%In other words, the eigenenergy spectrum before antisymmetrization does not have any meaning after antisymmetrization. 
%This means that the above identities do not say much about the degeneracy structure.  

To appreciate the meaning of this subtlety more concretely, it is important to note that the above trial states can be generated in the torus geometry as the zero-energy ground state of their respective model Hamiltonians, which exhibits a certain ground-state degeneracy specific to each state.
For example, the $Z_4$ parafermion state is generated by diagonalizing the fermionic 5-body $\delta$-function-interaction Hamiltonian, which has 15 degenerate zero-energy ground states.
Meanwhile, the model Hamiltonian generating the Halperin (330) state has 9 degenerate zero-energy ground states. 
Obviously, the ground-state degeneracies of these two model Hamiltonians do not match.

On the other hand, all the above identities in Eqs.~\eqref{eq:Z4_vs_A330}, \eqref{eq:Hf_vs_A551}, and \eqref{eq:Pf_vs_A331} are analytically proved by using the specific mathematical form of each trial wave function, which is written in the infinite planar geometry.
Logically speaking, such analytical proofs do not necessarily know about the ground-state degeneracy structure of the model Hamiltonian in the torus geometry. 
Thus, apparently, there is a dilemma between the analytical proof and the ground-state degeneracy mismatch in the torus geometry.

To resolve this dilemma, we conjecture that the above identities are to be interpreted in such a way that, in the torus geometry, only certain degenerate copies of the left-hand-side state in the bilayer mapping identities are exactly identical to the appropriate counterparts of the right-hand-side state.
It is explicitly shown in this work that the antisymmetrized Halperin (330) state has exactly the same ground-state degeneracy as the $Z_4$ parafermion state in a common momentum sector, where both Halperin (330) and $Z_4$ parafermion states occur as the ground states of their respective model Hamiltonians.
More importantly, in this momentum sector, the entire Hilbert space expanded by the degenerate copies of the antisymmetrized Halperin (330) state is exactly identical to that of the $Z_4$ parafermion state.
Based on this observation, we have a slightly stronger conjecture (at least for incompressible states) that, if there is a common momentum sector, where the two states related by the bilayer mapping identities have the same ground-state degeneracy, all the degenerate copies of the two states are exactly identical in such a momentum sector.

The situation is somewhat more complicated for the fermionic Haffnian state, whose model Hamiltonian has a diverging ground-state degeneracy as a function of particle number in the torus geometry~\cite{Hermanns11, Papic14}. 
On the other hand, the model Hamiltonian generating the Halperin (551) state has a finite ground-state degeneracy. 
So, it is impossible to reconcile these two vastly different ground-state degeneracies for the entire momentum sectors.
It would be interesting to check if our conjecture on the bilayer mapping identities works even in this situation.
To this end, a crucial question is if there is a common momentum sector, where the ground-state degeneracy of the fermionic Haffnian state is matched with that of the antisymmetrized Halperin (551) state. 
Unfortunately, at present, we are unable to perform a direct check to answer this question.
%Unfortunately, at present, the increasing ground-state degeneracy of the fermionic Haffnian state makes it difficult for us to perform a direct check to answer this question in the torus geometry.

Instead, in this work, we make use of the fact that the ground-state degeneracy issue does not occur in the spherical geometry, where both fermionic Haffnian and Halperin (551) states are non-degenerate.
Indeed, it is explicitly shown in this work that the overlap between the the antisymmetrized Halperin (551) and the fermionic Haffnian states is precisely unity in the spherical geometry, as predicted by Eq.~\eqref{eq:Hf_vs_A551}.
Actually, all the above trial states are also non-degenerate in the spherical geometry.
Consequently, the overlap between the antisymmetrized Halperin (330) and the $Z_4$ parafermion states is also precisely unity in the spherical geometry, as predicted by Eq.~\eqref{eq:Z4_vs_A330}.

Considering the numerical evidence obtained from both torus and spherical geometries, we think that it is reasonable to conjecture that, if there is a common momentum or angular momentum sector, where the ground-state degeneracies of the two states related by the bilayer mapping identities are matched, all the degenerate copies of the two states are exactly identical in such a momentum or angular momentum sector.
In fact, it turns out that such a momentum or angular momentum sector is usually where the uniform ground state occurs.

Under this understanding, for the sake of convenience, let us simply call the antisymmetrized Halperin (330) state the $Z_4$ parafermion state and the antisymmetrized Halperin (551) state the fermionic Haffnian state from this forward.
Intesretingly, it has been shown in a recent work~\cite{Repellin15} that the entire ground-state degeneracies of the $Z_k$ parafermion state can be fully reproduced by using the projective construction using a multilayer torus with twisted boundary conditions.

To summarize, the following four trial states are generated via the bilayer mapping;
(i) the Laughlin state, (ii) the antisymmetrized Halperin (551) state, which is identified as the fermionic Haffnian state under the proper understanding of the ground-state degeneracy explained above, (iii) the antisymmetrized product state of two CF seas at 1/6 filling, and (iv) the PH conjugate of the antisymmetrized Halperin (330) state, which is identified as the $Z_4$ parafermion state under the proper understanding of the ground-state degeneracy explained above.  
To investigate which trial state is most relevant at $\nu=7/3$, in this work, we compute the overlap between the exact $7/3$ ground state (which is the lowest energy state of the Coulomb interaction as a function of the Haldane pseudopotential variation) and the above four trial states by using ED up to $N=12$ in both torus and the spherical geometries.
As a result, it is shown that the PH conjugate of the $Z_4$ parafermion state has a substantial overlap with the exact 7/3 ground state around the Coulomb point, which is the highest among the four trial states.

The rest of this paper is organized as follows.
In Sec.~\ref{sec:Hamiltonian}, we provide the analytical expressions of the FQH Hamiltonians in both torus and spherical geometries, which are formulated in terms of the Haldane pseudopotentials.
In Sec.~\ref{sec:Trial_states}, we provide the concrete mathematical forms of the above four trial states and explain how to obtain them by applying the antisymmetrization and the PH conjugation operators onto the relevant bilayer quantum Hall states.   
In particular, it is proved analytically that the antisymmetrized projection of the Halperin (551) state is entirely equivalent to the fermionic Haffnian state.
In Sec.~\ref{sec:Overlap}, we provide the results for the overlap between the exact 7/3 ground state and the above four trial states. 
In Sec.~\ref{sec:Energy_spectrum}, we examine the energy spectrum, which exhibits an intriguing transition from the spectrum with the Laughlin-type magnetoroton structure to that with the specific quasi-degeneracy of the ground state, which is characteristic to the PH-conjugated $Z_4$ parafermion state. 
We conclude in Sec.~\ref{sec:Conclusion} with a summary of the results and a discussion on the future directions.

%%%%%%%%%%%%%
\section{Hamiltonian}
\label{sec:Hamiltonian}
%%%%%%%%%%%%%

In this section, we provide the analytical expressions of the FQH Hamiltonians in both torus and spherical geometries. 
Considering a recent experimental observation~\cite{Pan12}, here, we focus on the fully spin-polarized situation.
The goal of this section is to express the electron-electron interaction Hamiltonian in the Landau level with index $n$ ($n$LL) in terms of Haldane pseudopotentials~\cite{Haldane_Book}.
The pure Coulomb interaction can be obtained by choosing an appropriate set of Haldane pseudopotentials.

In the torus geometry~\cite{Yoshioka83,Haldane85}, the unit cell has a shape of parallelogram defined by two vectors ${\bf L}_1$ and ${\bf L}_2$ with the periodic boundary condition. 
The area of the unit cell is set equal to $|{\bf L}_1\times {\bf L}_2| = 2\pi l_B^2 N_{\phi}$, where $l_B$ is the magnetic length and $N_\phi$ is the number of flux quanta.
The aspect ratio of the unit cell is defined by $r_a=|{\bf L}_1|/|{\bf L}_2|$, which is set equal to unity in this work unless stated otherwise.
The $n$LL FQH Hamiltonian is written in terms of the torus basis states as follows:
\begin{align}
\label{eq:Torus_Hamiltonian}
H_{n\rm{LL}} = \frac{1}{2} \sum_{j_1, j_2, j_3, j_4} M_{j_1 j_2 j_3 j_4}^{(n)} c_{j_1}^{\dagger} c_{j_2}^{\dagger} c_{j_3} c_{j_4},
\end{align}
where ${c_j}^{\dagger}$ and ${c_j}$ are the creation and annihilation operators, respectively, acting on the $j$-th state with $j$ being the linear momentum quantum number.
The matrix element $M_{j_1 j_2 j_3 j_4}^{(n)}$ is given by~\cite{Haldane_Book,Jeong15}
\begin{align}
\label{eq:Torus_matrix_element}
M^{(n)}_{j_1 j_2 j_3 j_4}
=&\delta'_{j_1\!-j_4,t}\delta'_{j_1+j_2,j_3+j_4}
\nonumber \\
&\times \sum_{m=0}^\infty \frac{2V^{(n)}_m}{N_\phi} \sum_{\bf q} e^{iq_x(X_{j_1}-X_{j_3})}
 e^{-{q^2\over 2}} L_m\left(q^2\right),
\end{align}
where $X_j=2\pi j/|{\bf L}_2|$ for $j=1,2,\dots,N_{\phi}$ and ${\bf q}=s {\bf q}_1 + t {\bf q}_2$ $[s, t \in {\mathbb Z}]$ with ${\bf q}_1$ and ${\bf q}_2$ being the reciprocal vectors defined by the reciprocal relations, ${\bf L}_1\cdot {\bf q}_1=2\pi$ and ${\bf L}_2\cdot {\bf q}_2=2\pi$.
The primed Kronecker delta is defined as $\delta'_{i,j}=1$ if $i=j$ modulo $N_{\phi}$, and 0 otherwise.
$L_m(x)$ is the Laguerre polynomial. 
Called the Haldane pseudopotential, $V_m^{(n)}$ is the potential energy of an electron pair with relative angular momentum $m$ in the $n$LL.
For a given electron-electron interaction specified by its Fourier component $V_{\bf k}$, the Haldane pseudopotentials are obtained as follows~\cite{Haldane_Book}: 
\begin{align}
\label{eq:V_m}
V_{m}^{(n)} ={1 \over 2\pi } \int  d^2 {\bf k}~V_{\bf k} L_m(k^2) L_n^2\left({k^2\over 2}\right) e^{-k^2} .
\end{align}
Note that, in the case of the Coulomb interaction, the ${\bf q}=0$ component is excluded in the ${\bf q}$ summation in Eq.~\eqref{eq:Torus_matrix_element} to take into account the positive background correction.
It is convenient to vary the Coulomb interaction by adding the Haldane pseudopotential variations $\delta V^{(n)}_m$ to the pure Coulomb values $V_{{\rm Coul}, {m}}^{(n)}$.
In particular, we obtain the exact 7/3 ground state, $\Psi_{7/3}[\delta V_1^{(1)}]$, by diagonalizing the torus FQH Hamiltonian in Eq.~\eqref{eq:Torus_Hamiltonian} as a function of $\delta V^{(1)}_1$.
Note that all eigenstates of the torus FQH Hamiltonian can be classified in terms of the pseudomomentum ${\bf Q}=Q_1 {\bf q}_1 +Q_2 {\bf q_2} \equiv (Q_1, Q_2)$, which is conserved due to the translational invariance~\cite{Haldane85}.
%Here, $Q_1$ and $Q_2$ are integers between $0$ and ${\rm gcd}(N, N_{\phi})$~\cite{Haldane85}.
In the rectangular unit cell, $(Q_1,Q_2)= (Q_x, Q_y)$.

For the bilayer quantum Hall (BQH) system, the Hamiltonian is written as follows: 
\begin{align}
\label{eq:BQH_Hamiltonian}
H_{\rm BQH} = \frac{1}{2} \sum_{j_1, j_2, j_3, j_4} M_{j_1 j_2 j_3 j_4}^{ \sigma\sigma^\prime} c_{j_1\sigma}^{\dagger} c_{j_2\sigma^\prime}^{\dagger} c_{j_3\sigma^\prime} c_{j_4\sigma},
\end{align}
where the matrix element $M_{j_1 j_2 j_3 j_4}^{ \sigma\sigma^\prime}$ depends not only on the orbital momentum indices, $j_1,\cdots,j_4$, but also on the layer indices, $\sigma$ and $\sigma^\prime$. 
If $\sigma=\sigma^\prime$ ($\sigma \neq \sigma^\prime$), $M_{j_1 j_2 j_3 j_4}^{ \sigma\sigma^\prime}$ describes the intralayer (interlayer) interaction. 
Similar to Eq.~\eqref{eq:Torus_matrix_element}, the layer-dependent Haldane pseudopotentials $V^{\sigma\sigma^\prime}_m$ can be related with $M_{j_1 j_2 j_3 j_4}^{ \sigma\sigma^\prime}$  as follows:
 \begin{align}
\label{eq:BQH_matrix_element}
M^{\sigma\sigma^\prime}_{j_1 j_2 j_3 j_4}
=&\delta'_{j_1\!-j_4,t}\delta'_{j_1+j_2,j_3+j_4}
\nonumber \\
&\times \sum_{m=0}^\infty \frac{2V^{\sigma\sigma^\prime}_m}{N_\phi} \sum_{\bf q} e^{iq_x(X_{j_1}-X_{j_3})}
 e^{-{q^2\over 2}} L_m\left(q^2\right) ,
\end{align}  
where $V^{\sigma\sigma^\prime}_m = V^{\rm intra}_m$ and $V^{\rm inter}_m$ if $\sigma=\sigma^\prime$ and $\sigma \neq \sigma^\prime$, respectively.

So far, we have discussed the $n$LL FQH and the BQH Hamiltonians, which both have the two-body interaction between electrons.
As mentioned in Sec.~\ref{sec:Introduction}, for certain trial states, it is necessary to consider the model Hamiltonians with the $(k+1)$-body interaction.  
Specifically, the $Z_k$ parafermion state can be generated as the zero-energy ground state of the fermionic $(k+1)$-body $\delta$-function interaction Hamiltonian.  
See Appendix~\ref{appendix:k+1-body} for the details of the fermionic $(k+1)$-body $\delta$-function interaction Hamiltonian. 
Since we are particularly interested in the $Z_4$ parafermion state, in this work, we focus on the fermionic five-body $\delta$-function interaction Hamiltonian, whose concrete second quantization form in the torus geometry is given in Appendix~\ref{appendix:five-body}.

In the spherical geometry~\cite{Haldane83}, the $n$LL FQH Hamiltonian can be written for a two-body interaction $V({\bf r}_1, {\bf r}_2)$ as follows:
\begin{align}
\label{eq:Spherical_Hamiltonian}
H_{n\rm{LL}}=\frac{1}{2}\sum_{m_1 m_2 m_1^{\prime} m_2^{\prime}} \langle lm_1, lm_2 |V|lm_1^{\prime}, lm_2^{\prime} \rangle c_{m_1}^{\dagger}c_{m_2}^{\dagger}c_{m_2^{\prime}}c_{m_1^{\prime}},
\end{align}
where the orbital angular momentum $l$ is given by $l=Q+n$ for the $n$LL with $Q$ being the magnetic monopole strength. 
The azimuthal quantum numbers, $m_1$, $m_2$, $m_1^\prime$, and $m_2^\prime$, are summed over the range of  $\{-l, -l+1,\ldots, l-1, l\}$.	
An isotropic two-body interaction $V(r)$ can be represented in terms of the Haldane pseudopotentials as follows~\cite{Jain_Book}:
\begin{align}
&\langle lm_1, lm_2 |V(r)|lm_1^{\prime}, lm_2^{\prime} \rangle \nonumber \\
&=\sum_{L=0}^{2l}\sum_{M=-L}^{L} \langle lm_1,lm_2|LM\rangle V_L\langle LM|lm_1^{\prime},lm_2^{\prime}\rangle,
\end{align}
where the spherical Haldane pseudopotential $V_L$ is given as the potential energy of an electron pair with total angular momentum $L$, or equivalently with the relative angular momentum $2l-L$.
Specifically, for the pure Coulomb interaction, i.e., $V_{\rm Coul}({\bf r}_1, {\bf r}_2)=1/|{\bf r}_1-{\bf r}_2|$, 
\begin{align}
&\langle lm_1, lm_2 |V_{\rm Coul}|lm_1^{\prime},lm_2^{\prime} \rangle \nonumber\\
&=\frac{1}{R}\sum_{l^{\prime}}\sum_m\langle lm_1^{\prime},l^{\prime}m|lm_1\rangle\langle lm_2,l^{\prime}m|lm_2^{\prime}\rangle\langle lQ,l^{\prime}0|lQ\rangle ^2,
\end{align}
where the radius of the sphere $R$ is determined by $4\pi R^2 B=2Qhc/e$, or simply $R=\sqrt{Q}$ if we set the magnetic length $l_B=\sqrt{\hbar c/eB}$ equal to unity. 
The BQH Hamiltonian in the spherical geometry can be obtained by defining the layer-dependent spherical Haldane pseudopotentials similar to Eq.~\eqref{eq:BQH_Hamiltonian}.

%%%%%%%%%%%%%
\section{Trial states}
\label{sec:Trial_states}
%%%%%%%%%%%%%

As mentioned in Sec.~\ref{sec:Introduction}, we are interested in the following four trial states; (i) the Laughlin state, $\Psi_{333}$, (ii) the fermionic Haffnian state, $\Psi_{\rm Hf}$, which is shown to be equivalent to ${\cal A} \Psi_{551}$ under the proper understanding of the ground-state degeneracy in the torus geometry, (iii) the antisymmetrized product state of two CF seas at 1/6 filling, ${\cal A} \Psi_{^6{\rm CFS}\otimes ^6{\rm CFS}}$, and (iv) the PH conjugate of the $Z_4$ parafermion state, ${\cal C}_{\rm PH} \Psi_{Z_4}$, where $\Psi_{Z_4}$ is identified as ${\cal A} \Psi_{330}$ under the proper understanding of the ground-state degeneracy in the torus geometry.  
Below, we provide the concrete mathematical forms of the trial states and explain how to obtain them numerically by applying the antisymmetrization and the PH conjugation operators onto the relevant bilayer quantum Hall states.
See Ref.~\cite{Jeong15} for details on how to perform the antisymmetrization and the PH conjugation operators in second quantization.

\subsection{Laughlin state}
\label{sec:Laughlin}

The Laughlin state $\Psi_{333}$ is given as follows:
\begin{align}
\label{eq:333}
\Psi_{333} =  \prod_{i<j}^{N/2}\left(z_i-z_j\right)^3 \left(\omega_i-\omega_j\right)^3 \prod_{k,l}^{N/2} \left(z_k-\omega_l\right)^3 ,
\end{align}
which $z_i$ and $\omega_j$ denote the coordinates of the $i$-th and the $j$-th electron in each layer.
Evidently, $\Psi_{333}$ is invariant with respect to antisymmetrization as is.  
$\Psi_{333}$ can be obtained as the exact zero-energy ground state of the LLL FQH Hamiltonian in Eq.~\eqref{eq:Torus_Hamiltonian} with $V_1^{(0)}$ set equal to a nonzero positive number and all the other Haldane pseudopotentials to zero~\cite{Trugman85}.
Alternatively, the exact Coulomb ground state in the LLL can be used as an excellent approximation to $\Psi_{333}$.

\subsection{Fermionic Haffnian state}
\label{sec:Haffnian}

The fermionic Haffnian state is written as~\cite{Wen94}
\begin{align}
\label{eq:Haffnian}
\Psi_{\rm Hf} =  \Phi_1^{3} {\rm Hf}\left(1\over \left(Z_i -Z_j\right)^2\right) =\Phi_1^{3}{\rm Det}\left(1\over Z_i -Z_j \right) ,
\end{align}
where $\Phi_1 = \prod_{i<j}^N (Z_i-Z_j)$ with $Z_i$ being the unified coordinates defined as $Z_i= z_i$ and $Z_{i+N/2}=\omega_i$ with $i=1, 2, \dots, N/2$.
Note that $z_i$ and $w_i$ denote the coordinates of electrons in the top and bottom layers, respectively, while there is no physical distinction between two different layer degrees of freedom at the final wave function, which is totally antisymmetried as a whole.
$\Phi_1^3$ is equivalent to $\Psi_{333}$.
${\rm Hf}$ denotes the Haffnian of a symmetric matrix, which is related with the determinant via ${\rm Hf}(M_{ij}^2) = {\rm Det}(M_{ij})$~\cite{Greiter92}.
Meanwhile, the Halperin (551) state $\Psi_{\rm 551}$~\cite{Halperin83,Yoshioka89} is written as 
\begin{align}
\label{eq:551}
\Psi_{\rm 551} = \prod_{i<j}^{N/2}\left(z_i-z_j\right)^5 \left(\omega_i-\omega_j\right)^5 \prod_{k,l}^{N/2}\left(z_k-\omega_l\right).
\end{align}

Now, we would like to prove that the antisymmetrized projection of the Halperin (551) state, ${\cal A} \Psi_{\rm 551}$, is entirely equivalent to $\Psi_{\rm Hf}$ up to a normalization constant.
To this end, it is convenient to rewrite ${\cal A} \Psi_{\rm 551}$ as follows:
\begin{align}
\label{eq:A(551)}
{\cal A}\Psi_{551} = \Phi_1 {\cal S}\prod_{i<j}^{N/2}\left(z_i-z_j\right)^4\left(\omega_i-\omega_j\right)^4,
\end{align}
where ${\cal S}$ is the symmetrization operator. 
In order to express Eq.~\eqref{eq:A(551)} in the form of a paired state, we exploit two identities for the symmetrized Jastrow factor. 
The first identity is given by
\begin{align}
\label{eq:symmetric_polynomial_1}
&{\cal S}\prod^{N/2}_{i<j}(z_i-z_j)^4(\omega_i-\omega_j)^4
\nonumber \\
&= 2^{\left(1-N/2\right)} 
\left[{\cal S}\prod^{N/2}_{i<j}(z_i-z_j)^2(\omega_i-\omega_j)^2\right]^2, 
\end{align}
which is proved in Appendix~\ref{appendix:identity}. 
The second identity is the well-known analytical relationship between the symmetrized Halperin (220) and the bosonic MR Pfaffian wave function, which is fundamentally due to Cauchy's identity~\cite{Greiter92_Paired}: 
\begin{align}
\label{eq:symmetric_polynomial_2}
{\cal S} \prod_{i<j}^{N/2}\left(z_i-z_j\right)^2\left(\omega_i-\omega_j\right)^2 
= C_{N/2} \Phi_1 {\rm Pf}\left({1\over Z_i-Z_j}\right),
\end{align}
where the constant factor $C_n=(-1)^{n(n-1)/2} n!$. 
Above, ${\rm Pf}$ denotes the Pfaffian of a skew-symmetric matrix.
Using the two identities in Eqs.~\eqref{eq:symmetric_polynomial_1} and \eqref{eq:symmetric_polynomial_2},
one can rewrite ${\cal A}\Psi_{551}$ up to a normalization constant as follows:
\begin{align}
\label{eq:Equivalence_A(551)_Hf}
{\cal A} \Psi_{551} = \Phi_1^3 \left[ {\rm Pf} \left({1\over Z_i - Z_j}\right) \right]^2= \Phi_1^3 {\rm Det} \left({1\over Z_i - Z_j}\right) ,
\end{align}
where it is used that $[{\rm Pf}(M_{ij})]^2={\rm Det} M_{ij}$.
This proves that ${\cal A} \Psi_{\rm 551}$ is entirely equivalent to $\Psi_{\rm Hf}$ up to a normalization constant.

It is interesting to mention that the bosonic counterpart of Eq.~\eqref{eq:Equivalence_A(551)_Hf}, i.e., the symmetrized product of two Laughlin states at quarter filling is equivalent to the bosonic Haffnian state at half filing, has been considered previously. 
Mathematically, the bosonic counterpart of Eq.~\eqref{eq:Equivalence_A(551)_Hf} can be written as
\begin{align}
{\cal S} \Psi_{440} = \Phi_1^2 {\rm Hf}\left(1\over \left(Z_i -Z_j\right)^2\right) = \Phi_1^2 {\rm Det}\left(1\over Z_i -Z_j \right),
\end{align} 
which was mentioned before in Ref.~\cite{Ardonne11}, while no proof was provided there.
Also, the above relationship was previously discussed in the context of the pattern-of-zeros and vertex algebra approaches~\cite{Barkeshli11}.

According to Eq.~\eqref{eq:Equivalence_A(551)_Hf}, one can generate the fermionic Haffnian state as the antisymmetrized projection of the Halperin (551) state. 
Similar to the Laughlin state, the Halperin (551) state can be obtained as the exact zero-energy ground state of the BQH Hamiltonian in Eq.~\eqref{eq:BQH_Hamiltonian} with $V^{\rm intra}_{1,3}$ and $V^{\rm inter}_0$ set equal to nonzero positive numbers and all the other Haldane pseudopotentials to  zero~\cite{Yoshioka89}. 
%In Sec.~\ref{sec:Overlap}, we show the overlap between the exact 7/3 ground state and the antisymmetrized projection of the Halperin (551) state obtained in the same momentum sector, where the exact 7/3 ground state belongs. 
%Obviously, if the Halperin (551) state occurs in a different momentum sector, the overlap becomes strictly zero. 

It is worthwhile to mention that the antisymmetrized projection of the bilayer state is not necessarily incompressible even if the original bilayer state is so.
The Halperin (551) state is incompressible, whereas the Haffnian state is known to be not~\cite{Read09}.
This result is based on the conformal field theory with conformal blocks, which correspond to trial wave functions.
It is shown that the conformal field theory becomes irrational if its conformal block corresponds to the Haffnian wave function. 
As a result, the number of excitation types in the Haffnian state is not finite.
This property manifests itself in finite-size numerical calculations as a diverging degeneracy of the ground state in the torus geometry and that of the quasihole state in the spherical geometry as a function of particle number.
Both numerical and analytical studies have shown that this is indeed the case for the bosonic Haffnian state~\cite{Hermanns11,Papic14}. 
We believe that the same should be true for the fermionic Haffnian state.

For this reason, in the torus geometry, there is a serious issue of the ground-state degeneracy mismatch between the fermionic Haffnian state, which has a diverging ground-state degeneracy, and the antisymmetrized Halperin (551) states, which has a finite ground-state degeneracy. 
The Halperin (551) state has the 24-fold degeneracy in the torus geometry, which can be understood in terms of its root configurations.
Specifically, we have found that there are four distinct root configurations for the Halperin (551) state:
(i) $|XX\circ\circ\circ\circ\cdots\rangle$, (ii) $|\uparrow\circ\downarrow\circ\circ\circ\cdots\rangle$, (iii) $|\uparrow\circ\circ\downarrow\circ\circ\cdots\rangle$, and (iv) $|\uparrow\circ\circ\circ\downarrow\circ\cdots\rangle$, where $\uparrow$ and $\downarrow$ indicate a site occupied by up and down spins, respectively, $\circ$ indicates an empty site, $XX=\uparrow\downarrow+\downarrow\uparrow$, and the ellipsis denotes the repetition of a given root configuration. 
Each of these four root configurations has a six-fold degeneracy via the center-of-mass shift, altogether generating 24 degenerate states.   
Note that the root configurations of the Halperin (551) state can be constructed similarly to those of the Halperin (331) state~\cite{Seidel08}.

After antisymmetrization, the root configuration in (i) becomes $|\bullet\bullet\circ\circ\circ\circ\cdots\rangle$, where $\bullet$ indicate a filled site.
Meanwhile, the root configurations in both (ii) and (iv) reduce to a single identical state, $|\bullet\circ\bullet\circ\circ\circ\cdots\rangle$.
Each of these two antisymmetrized root configurations, $|\bullet\bullet\circ\circ\circ\circ\cdots\rangle$ and $|\bullet\circ\bullet\circ\circ\circ\cdots\rangle$, generates six degenerate copies via the center-of-mass shift.
On the other hand, the root configuration in (iii) becomes $|\bullet\circ\circ\bullet\circ\circ\cdots\rangle$, which has three degenerate copies since the period of the center-of-mass shift is halved.
As a consequence, the antisymmetrized Halperin (551) state has altogether the 15-fold ground-state degeneracy in the torus geometry. 

More importantly, it can be shown by counting the pseudomomentum of each root configuration that there are two degenerate copies of the antisymmetrized Halperin (551) state at ${\bf Q}=(N/2,N/2)$, where the exact 7/3 ground state occurs.
As mentioned in Sec.~\ref{sec:Introduction}, however, it is unclear at present whether the ground-state degeneracy of the antisymmetrized Halperin (551) state is the same as that of the fermionic Haffnian state in this momentum sector.

Fortunately, the ground-state degeneracy issue does not occur in the spherical geometry, where both fermionic Haffnian and Halperin (551) states are non-degenerate.  
Specifically, in the spherical geometry, the fermionic Haffnian state can be directly obtained as the exact, non-degenerate, zero-energy ground state of the following three-body interaction Hamiltonian~\cite{Green02}:
\begin{align}
\label{eq:Haffnian_Hamiltonian}
H_{\rm Hf} = &\sum_{i\neq j\neq k} \Big[ V_0P_{ijk}\left(3N_{\phi}/2-3\right)
\nonumber \\
&+V_2P_{ijk}\left(3N_{\phi}/2-5\right)+V_3P_{ijk}\left(3N_{\phi}/2-6\right) \Big],
\end{align}
where $P_{ijk}(L)$ is the projection operator onto the three-particle state at a given total angular momentum $L$. 
The $V_1$ term is absent due to a symmetry reason~\cite{Simon07}.

We have confirmed that, in the spherical geometry, the ground state of $H_{\rm Hf}$ is indeed exactly identical to the antisymmetrized projection of the Halperin (551) state up to machine accuracy. 
Based on this confirmation, as far as the fermionic Haffnian state is concerned, we put more emphasis on the results obtained in the spherical geometry. 
In Sec.~\ref{sec:spherical_geometry}, we show the results for the overlap between the exact 7/3 ground and the antisymmetrized Halperin (551) states in the spherical geometry.
It turns out, fortunately, that the overlap results obtained in the spherical geometry are overall consistent with those in the torus geometry, as reported in Sec.~\ref{sec:Overlap}.

\subsection{Antisymmetrized product state of two CF seas at 1/6 filling}
\label{sec:2CFS}

Here, we consider the antisymmetrized product state of two CF seas at 1/6 filling, ${\cal A} \Psi_{^6 {\rm CFS} \otimes ^6 {\rm CFS}}$.
To this end, it is necessary to know how to obtain the CF sea state at 1/6 filling, $\Psi_{^6 {\rm CFS}}$.
Na\"{i}vely, one may guess that $\Psi_{^6 {\rm CFS}}$ is obtained as the ground state of the Coulomb interaction at $\nu=1/6$ in the LLL.
Unfortunately, however, this guess is not correct since the actual ground state is likely to be the Wigner crystal of composite fermions rather than the quantum Hall liquid state~\cite{Archer13, Rhim15}.
In the torus geometry, this manifests itself as the fact that the ground state occurs at odd pseudomomenta, indicating that it is not a uniform state.  
In this sense, the previously-mentioned phase diagram in Fig.~\ref{fig:bilayer_ground_state}~(a) is in fact inaccurate. 
The actual 1/3 bilayer quantum Hall ground state at $d/l_B \gg 1$ is predicted to be the product state of two Wigner crystals of composite fermions at 1/6 filling.

For this reason, we instead construct $\Psi_{^6 {\rm CFS}}$ as the exact ground state of the LLL FQH Hamiltonian in Eq.~\eqref{eq:Torus_Hamiltonian} with $V^{(0)}_{1,3,5}$ set equal to unity and all the other Haldane pseudopotentials zero. 
It is important to note that ${\cal A} \Psi_{^6 {\rm CFS} \otimes ^6 {\rm CFS}}$ does not necessarily describe a compressible phase even if $\Psi_{^6 {\rm CFS} \otimes ^6 {\rm CFS}}$ is compressible. 
The reason is in some sense similar to why the antisymmetrized incompressible state is not necessarily incompressible, as shown in the example of $\Psi_{\rm Hf}={\cal A} \Psi_{551}$.
Perhaps, a more directly relevant example is the previously-mentioned antisymmetrized product state of two CF seas at quarter filling, ${\cal A}\Psi_{^4{\rm CFS} \otimes ^4{\rm CFS}}$, which has a significant square of overlap (over $90\%$ when PH-symmetrized in the $N=12$ system) with the exact $5/2$ ground state at the Coulomb point~\cite{Jeong15}, which is known to be incompressible.
%Meanwhile, the simple product state of two CF seas at quarter filling, $\Psi_{^4{\rm CFS} \otimes ^4{\rm CFS}}$, is apparently compressible.
It is not clear at this point whether ${\cal A} \Psi_{^6 {\rm CFS} \otimes ^6 {\rm CFS}}$ is indeed incompressible.
Despite this uncertainty, we believe that it is worthwhile to investigate how large overlap it can have with the exact 7/3 ground state around the Coulomb point. 
In some sense, the overlap can provide us with a hint for the compressibility of ${\cal A} \Psi_{^6 {\rm CFS} \otimes ^6 {\rm CFS}}$.

\subsection{PH-conjugated $Z_4$ parafermion state}
\label{sec:C_PH_Z_4}

Now, let us consider the PH-conjugated $Z_4$ parafermion state, ${\cal C}_{\rm PH} \Psi_{Z_4}$.
As mentioned previously, the $Z_4$ parafermion state can be obtained as the antisymmetrized projection of the Halperin (330) state~\cite{Rezayi10, Barkeshli10} with the Halperin (330) state defined as
\begin{align}
\label{eq:330} 
\Psi_{\rm 330} = \prod_{i<j}^{N/2}\left(z_i-z_j\right)^3 \left(\omega_i-\omega_j\right)^3, %\prod_{k,l}^{N/2}\left(z_k-\omega_l\right)^0.
\end{align}
which can be in turn obtained as the exact zero-energy ground state of the BQH Hamiltonian in Eq.~\eqref{eq:BQH_Hamiltonian} with $V^{\rm intra}_{1}$ set equal to a nonzero positive number and all the other Haldane pseudopotentials to zero.  
Practically, we use the fact that $\Psi_{330}$ is the direct product of two Laughlin states, each of which can be obtained in a much smaller Hilbert space.  
For example, for $N=12$ in the torus geometry, the size of the Hilbert space for the 2/3 BQH system is around $3.6 \times 10^{15}$, whereas that for the 1/3 FQH system is around $2.9 \times 10^{6}$.
Once $\Psi_{\rm 330}$ is obtained, we apply the antisymmetrization operator to obtain ${\cal A} \Psi_{\rm 330}=\Psi_{Z_4}$ and then the PH conjugation operator to obtain ${\cal C}_{\rm PH} \Psi_{Z_4}$.

As mentioned in Sec.~\ref{sec:Introduction}, there is an issue of the ground-state degeneracy mismatch between the antisymmetrized Halperin (330) and the $Z_4$ parafermion states in the torus geometry.
Fortunately, this ground-state degeneracy issue is much less severe, compared to that between the antisymmetrized Halperin (551) and the fermionic Haffnian states discussed in the preceding section.
Specifically, there is a certain momentum sector in the torus geometry, where all degenerate copies of the antisymmetrized Halperin (330) state are exactly identical to the appropriate counterparts of the $Z_4$ parafermion state.

To understand what this momentum sector is, let us consider the root configurations of the Halperin (330) and the $Z_4$ parafermion states.
First, the Halperin (330) state is constructed as the direct product between two Laughlin states in the top ($\uparrow$) and bottom ($\downarrow$) layers, each of which has three degenerate copies via the center-of-mass shift, leading to the nine-fold ground-state degeneracy of the Halperin (330) state~\cite{Keski-Vakkuri93}.
Specifically, the Laughlin state has the following three root configurations: $| \bullet \circ \circ \bullet \circ \circ \bullet \circ \circ \cdots \rangle$, $| \circ \bullet \circ \circ \bullet \circ \circ \bullet \circ \cdots \rangle$, and $| \circ \circ \bullet \circ \circ \bullet \circ \circ \bullet \cdots \rangle$, where $\bullet$ and $\circ$ indicate filled and empty sites, respectively. 
These three root configurations are related with each other via the center-of-mass shift. 
Evaluating the pseudomomentum of each root configuration, one can show that three degenerate copies of the Laughlin state occur at $Q_y=0$, $N_\phi /3$, and $2 N_\phi /3$ for odd, and $N_\phi /6$, $N_\phi /2$, and $5 N_\phi /6$ for even particle numbers, where $N_\phi$ is the number of flux quanta.  
We set $N_\phi=3 N_\uparrow=3 N_\downarrow$ since we are interested in the BQH system with the top and bottom layers having the (equal) particle number $N_\uparrow$ and $N_\downarrow$, respectively. 
Note that there is no flux shift in the torus geometry.
%The above momenta can be derived by considering the root configurations of the Laughlin state; $| \bullet \circ \circ \bullet \circ \circ \bullet \circ \circ \cdots \rangle$, $| \circ \bullet \circ \circ \bullet \circ \circ \bullet \circ \cdots \rangle$, and $| \circ \circ \bullet \circ \circ \bullet \circ \circ \bullet \cdots \rangle$, where $\bullet$ and $\circ$ indicate filled and empty sites, respectively. 

The root configurations of the Halperin (330) state can be obtained by taking the direct product between those of the two Laughlin states in the top and bottom layers.
Among the nine degenerate copies obtained from this product, the Halperin (330) state can occur particularly at $Q_y=Q_{y\uparrow}+Q_{y\downarrow}=0$ (mod $N_\phi$) by combining $(Q_{y\uparrow}, Q_{y\downarrow})= (0,0)$, $(N_\phi /3, 2 N_\phi /3)$, and $(2 N_\phi /3, N_\phi /3)$ for odd $N_\uparrow$ and $N_\downarrow$, and $(N_\phi /2, N_\phi /2)$, $(N_\phi /6, 5 N_\phi /6)$, and $(5 N_\phi /6, N_\phi /6)$ for even $N_\uparrow$ and $N_\downarrow$.
As a consequence, the momentum sector at $Q_y=0$ (mod $N_\phi$) has the triple degeneracy. 
Also, due to the center-of-mass shift, there are other similar triple degeneracies in the momentum sectors at $Q_y=N$ and $2N$ (mod $N_\phi$) with $N=N_\uparrow+N_\downarrow$. 
It is worthwhile to mention that, along the $x$-direction, the momentum of all the above degenerate states is the same; $Q_x=Q_{x\uparrow}+Q_{x\downarrow}=0$ (mod $N^\prime$) with $N^\prime= \textrm{gcd}(N,N_\phi)$.

For odd $N_\uparrow$ and  $N_\downarrow$, two degenerate copies of the Halperin (330) state can be obtained from $(Q_{y\uparrow},Q_{y\downarrow})=(N_\phi /3, 2 N_\phi /3)$ and $(2 N_\phi /3, N_\phi /3)$, both of which reduce to a single identical state at $Q_y=0$ (mod $N_\phi$) after antisymmetrization.
%the two antisymmetrized projections of the Halperin (330) state obtained from $(Q_{y\uparrow},Q_{y\downarrow})=(N_\phi /3, 2 N_\phi /3)$ and $(2 N_\phi /3, N_\phi /3)$, respectively, reduce to a single identical state at $Q_y=0$ (mod $N_\phi$).
Meanwhile, the antisymmetrized projection of the Halperin (330) state obtained from $(Q_{y\uparrow}, Q_{y\downarrow})= (0,0)$ remains distinct from this state.
Consequently, a double ground-state degeneracy is generated at $Q_y=0$.
Since $Q_x=0$ for all the above states, this means that there is the double ground-state degeneracy for the antisymmetrized Halperin (330) state at ${\bf Q}=(Q_x,Q_y)=(0,0)$. 
Similarly, for even $N_\uparrow$ and $N_\downarrow$, two degenerate copies of the Halperin (330) state can be obtained from $(Q_{y\uparrow},Q_{y\downarrow})=(N_\phi /6, 5 N_\phi /6)$ and $(5 N_\phi /6, N_\phi /6)$, both of which reduce to a single identical state at $Q_y=0$ (mod $N_\phi$) after antisymmetrization. 
Also, a different single state is obtained by antisymmetrizing the Halperin (330) state at $(Q_{y\uparrow}, Q_{y\downarrow})=(N_\phi /2, N_\phi /2)$, consequently generating the double ground-state degeneracy at ${\bf Q}=(0,0)$ similar to the case of odd $N_\uparrow$ and  $N_\downarrow$.
Eventually, there is altogether the six-fold ground-state degeneracy due to the center-of-mass shift.

It is important to note that, strictly speaking, antisymmetrization would annihilate any root configurations obtained in the momentum sector satisfying $Q_{y\uparrow}=Q_{y\downarrow}$, i.e., where electrons with opposite layer indices are exactly on top of each other. 
Fortunately, there are extra ``fluctuation configurations,'' which are derived from the root configurations away from the thin-torus limit. 
These fluctuation configurations can generally survive antisymmetrization.  
%In summary, after antisymmetrization of the Halperin (330) state, there is a double ground-state degeneracy in the ${\bf Q}=(0,0)$ (mod $N^\prime$) momentum sector. 

Now, let us switch gears and consider the ground-state degeneracy of the $Z_4$ parafermion state.
There are altogether fifteen different root configurations for the $Z_4$ parafermion state:
(i) $| \bullet \bullet \bullet \bullet \circ \circ \cdots \rangle$ and its six other center-of-mass-shifted copies,
(ii) $| \bullet \bullet \bullet \circ \bullet \circ \cdots \rangle$ and its six other center-of-mass-shifted copies, and
(iii) $| \bullet \bullet \circ \bullet \bullet \circ \cdots \rangle$ and its three other center-of-mass-shifted copies.
Physically, these root configurations can be understood as all possible ways of distributing four electrons within the unit cell of six sites, followed by the center-of-mass shift~\cite{Ardonne08,Liu15}. 
Having absolutely zero amplitudes of five-electron cluster, these root configurations generate the zero-energy ground states of the fermionic five-body $\delta$-function interaction Hamiltonian, which imposes an energy cost only on five-electron cluster.  
Note that, by definition, the $Z_k$ parafermion state is the zero-energy ground state of the fermionic $(k+1)$-body $\delta$-function interaction Hamiltonian.
See Appendix~\ref{appendix:k+1-body} for the details of the fermionic $(k+1)$-body $\delta$-function interaction Hamiltonian.

Counting the pseudomomenta allowed for all the above root configurations, one can show that there is the double ground-state degeneracy for the $Z_4$ parafermion state at ${\bf Q}=(0,0)$ (mod $N^\prime$). 
%This conclusion is exactly the same as that in the case of the antisymmetrized Halperin (330) state. 
Therefore, the ground-state degeneracy of the $Z_4$ parafermion state is exactly matched with that of the antisymmetrized Halperin (330) state in this momentum sector.

Now, we would like to show that, in this momentum sector, the two degenerate copies of the $Z_4$ parafermion state are indeed exactly identical to those of the antisymmetrized Halperin (330) state.  
To this end, we obtain the $Z_4$ parafermion state by directly diagonalizing the fermionic five-body $\delta$-function interaction Hamiltonian. 
See Appendix~\ref{appendix:five-body} for the concrete second quantization form of the fermionic five-body $\delta$-function interaction Hamiltonian in the torus geometry.

Specifically, we perform ED of the fermionic five-body $\delta$-function interaction Hamiltonian to obtain the $Z_4$ parafermion state in the finite-size system with $N=8$ and $N_\phi=12$. 
In the mean time, the antisymmetrized Halperin (330) state is obtained in the same finite-size system by making the product of two Laughlin states, each of which is obtained in the finite-size system with $N=4$ and $N_\phi=12$ as the exact zero-energy ground state of the LLL FQH Hamiltonian with $V_1^{(0)}$ set equal to a nonzero positive number and all the other Haldane pseudopotentials to zero. 
In the end, we compute the overlap between the two degenerate copies of the $Z_4$ parafermion state and those of the antisymmetrized Halperin (330) state. 
As a result, it is explicitly shown that the two degenerate copies of the antisymmetrized Halperin (330) state at ${\bf Q}=(0,0)$ have exactly the unity overlap with their respective counterparts of the $Z_4$ parafermion state. 
In other words, at ${\bf Q}=(0,0)$, all degenerate copies of the $Z_4$ parafermion state can be obtained by antisymmetrizing those of the Halperin (330) state.

As mentioned previously, the issue of the ground-state degeneracy mismatch does not occur in the spherical geometry.
To explicitly confirm this to be the case, first, we have directly obtained the $Z_4$ parafermion state via the Jack polynomial~\cite{Bernevig08}.  
As before, the Halperin (330) state is obtained by making the product of two Laughlin states in the spherical geometry. 
It is explicitly shown that the $Z_4$ parafermion state obtained via the Jack polynomial is indeed exactly identical to the antisymmetrized Halperin (330) state, both occurring at the $L=0$ sector without any degeneracy.

It is worthwhile to mention that a recent numerical work by Peterson {\it et al.} in the spherical geometry~\cite{Peterson15} has shown that $\Psi_{Z_4}$ has a significant overlap with exact $8/3$ state with full spin polarization. 
In the limit of zero Landau level mixing, the same significant overlap should be obtained between ${\cal C}_{\rm PH} \Psi_{Z_4}$ and the exact 7/3 ground state.
We have explicitly confirmed this to be the case by performing ED in the spherical geometry.  
In fact, we have been able to perform ED in a much larger system than those studied by Peterson {\it et al.}, who studied up to the $N=16$ system at $\nu=8/3$, which corresponds to the $N=6$ system at $\nu=7/3$ after PH conjugation. 
As one can see in Fig.~\ref{fig:overlap_sphere}, we report ED results at $\nu=7/3$ up to $N=10$ for ${\cal C}_{\rm PH} \Psi_{Z_4}$, while up to $N=12$ for both $\Psi_{333}$ and $\Psi_{\rm Hf}$.

\subsection{Other related bilayer quantum Hall states}

In connection with the $Z_4$ parafermion state, there are several other related non-Abelian trial states in the bilayer quantum Hall system at $\nu=n+2/3$ with $n$ being the filled Landau level index. 
These trial states include the intralayer Pfaffian state~\cite{Ardonne02}, the interlayer Pfaffian state~\cite{Barkeshli10_Classification,Geraedts15}, the Fibonacci state~\cite{Wen91, Vaezi14}, and the Bonderson-Slingerland state~\cite{Bonderson08}.

We remark that it is not easy to construct these trial states in the torus geometry since the parent Hamiltonian generating each state as the zero-energy ground state is unknown. 
On the other hand, in the spherical geometry, one can obtain the amplitudes of each state in second quantization basis by making use of the Jack polynomial representation~\cite{Bernevig09}.
While these trial states could be important in some parameter regimes of the bilayer quantum Hall system, in this work, we are only interested in the single-layer quantum Hall states, which can be obtained by antisymmetrizing the simple Abelian bilayer quantum Hall states such as the Halperin (330) and (551) states.

%%%%%%%%%%%%%
\section{Results}
\label{sec:Results}
%%%%%%%%%%%%%

%%%%%%%%%%%%%
\subsection{Overlap}
\label{sec:Overlap}
%%%%%%%%%%%%%

In this section, we compute the overlap between the exact 7/3 ground state and each of the four trial states, $\Psi_{333}$, $\Psi_{\rm Hf}$, ${\cal A} \Psi_{^6{\rm CFS}\otimes ^6{\rm CFS}}$, and  ${\cal C}_{\rm PH} \Psi_{Z_4}$ by using ED in both torus and spherical geometries up to $N=12$.
In the preceding section, we have discussed how to obtain the four trial states in both torus and spherical geometries.
It is emphasized that the ground-state degeneracy of each trial state in the torus geometry should be properly understood as discussed in the preceding section.

Note that we obtain the exact 7/3 ground state in the torus geometry by exactly diagonalizing the SLL FQH Hamiltonian in Eq.~\eqref{eq:Torus_Hamiltonian} as a function of the Haldane pseudopotential variation $\delta V_1^{(1)}/V_1^{(1)}$ with $V_1^{(1)}=0.415419$.
To emphasize its dependence on $\delta V^{(1)}_1$, the exact 7/3 ground state is denoted as $\Psi_{7/3}[\delta V_1^{(1)}]$ in the torus geometry.
Meanwhile, in the spherical geometry, we focus on the exact 7/3 ground state only at the pure Coulomb interaction.

Before presenting the overlap results, it is worthwhile to explain our choice of particle numbers. 
While $\Psi_{\rm Hf}$ and ${\cal C}_{\rm PH} \Psi_{Z_4}$ can be constructed in principle at any particle numbers by using appropriate many-body model Hamiltonians, it is convenient to choose even particle numbers in our bilayer mapping approach. 
Also, due to the property of the antisymmetrized product state, ${\cal A} \Psi_{^6{\rm CFS}\otimes ^6{\rm CFS}}$ can be constructed only when the particle number is a multiple of four~\cite{Jeong15}.

%%%%%%%%%%%%%%%%%%%%%%%%%%%%%%%%%%%%%%%%%%%%%%%%%%%%%%%%%%%%%%%%%%%%%%%%
\begin{figure*}[t!]
\includegraphics[width=1\textwidth]{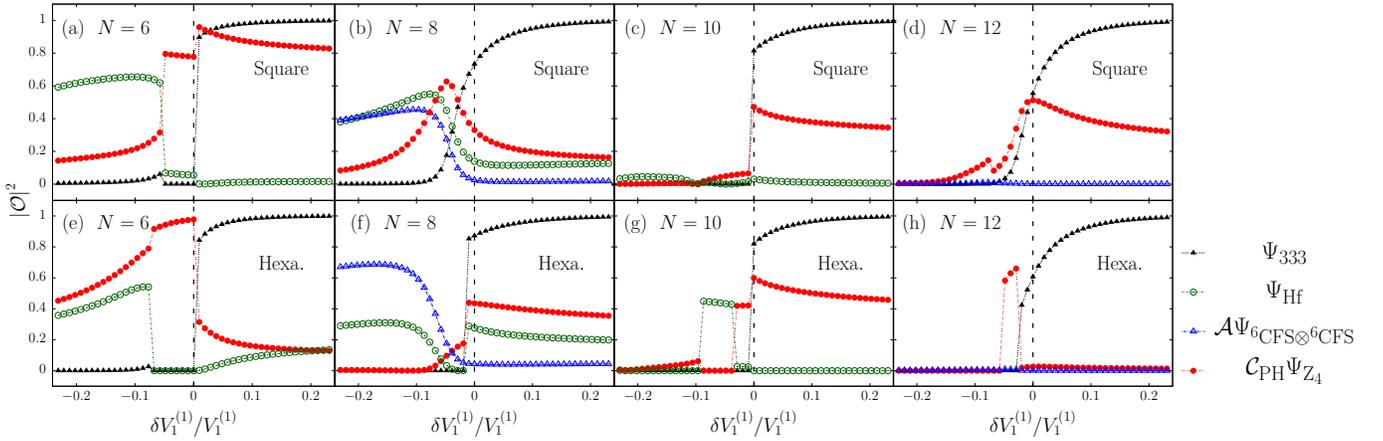}
\caption{
(Color online) Square of overlap $|{\cal O}|^2$ between the exact 7/3 ground state, $\Psi_{7/3}[\delta V_1^{(1)}]$, and each of the four trial states, $\Psi_{333}$, $\Psi_{\rm Hf}$, ${\cal A} \Psi_{^6{\rm CFS}\otimes ^6{\rm CFS}}$, and  ${\cal C}_{\rm PH} \Psi_{Z_4}$, as a function of the Haldane pseudopotential variation $\delta V_1^{(1)}/V_1^{(1)}$ with $V_1^{(1)}=0.415419$.
The exact 7/3 ground state, $\Psi_{7/3}[\delta V_1^{(1)}]$, is obtained as the exact lowest energy state of the Coulomb interaction Hamiltonian with the Haldane pseudopotential variation $\delta V_1^{(1)}$ at pseudomomentum ${\bf Q}=(N/2, N/2)$. 
Note that the lowest energy state always occurs at ${\bf Q}=(N/2, N/2)$ for the entire range of $\delta V_1^{(1)}$. 
The shape of the unit cell is square in (a)\mbox{--}(d), and hexagonal in (e)\mbox{--}(h). 
The electron-electron interaction is purely Coulombic at $\delta V_1^{(1)}=0$. 
}
\label{fig:overlap}
\end{figure*}
%%%%%%%%%%%%%%%%%%%%%%%%%%%%%%%%%%%%%%%%%%%%%%%%%%%%%%%%%%%%%%%%%%%%%%%%

\subsubsection{Torus geometry}
\label{sec:torus_geometry}

Figure~\ref{fig:overlap} shows the square of overlap between $\Psi_{7/3}[\delta V_1^{(1)}]$ and each of the four trial states, $\Psi_{333}$, $\Psi_{\rm Hf}$, ${\cal A} \Psi_{^6{\rm CFS}\otimes ^6{\rm CFS}}$, and  ${\cal C}_{\rm PH} \Psi_{Z_4}$ as a function of the Haldane pseudopotential variation $\delta V_1^{(1)}/V_1^{(1)}$ for various particle numbers in the torus geometry.
Note that the torus geometry can accommodate different parallelogram shapes for the unit cell by continuously varying the angle between two lateral vectors ${\bf L}_1$ and ${\bf L}_2$, which can deform the unit cell from square to hexagon~\cite{Papic12}.
As one can see, there are somewhat wild fluctuations in the behavior of the overlap across various particle numbers. 
Despite these fluctuations, it is possible to extract the following overall behaviors of the overlap.

First, the overlap between $\Psi_{7/3}[\delta V_1^{(1)}]$ and $\Psi_{333}$ shows the most stable behavior as a function of $\delta V_1^{(1)}/V_1^{(1)}$ regardless of particle number.
Specifically, the overlap is close to unity for sufficiently large positive $\delta V_1^{(1)}/V_1^{(1)}$, but decreases fast as $\delta V_1^{(1)}/V_1^{(1)}$ approaches the Coulomb point.
At moderately negative $\delta V_1^{(1)}/V_1^{(1)}$, the overlap becomes negligibly small.
In contrast to the situation in the LLL, $\Psi_{333}$ has a rather weak chance of representing the exact Coulomb ground state in the SLL.

Second, as for the fermionic Haffnian state, it is important to properly define the overlap between $\Psi_{7/3}[\delta V_1^{(1)}]$ and $\Psi_{\rm Hf}$.
As mentioned in Sec.~\ref{sec:Haffnian}, $\Psi_{\rm Hf}$ has a multiple ground-stare degeneracy in the torus geometry, which diverges as a function of particle number~\cite{Lee15}.
Meanwhile, $\Psi_{7/3}[\delta V_1^{(1)}]$ is uniquely obtained as the lowest energy state of the Coulomb interaction, which generally does not have any exact ground-state degeneracy (except that due to the center-of-mass shift). 
This leads to an ambiguity regarding the definition of the overlap between $\Psi_{7/3}[\delta V_1^{(1)}]$ and $\Psi_{\rm Hf}$.

Mathematically, one needs to properly define the overlap between a non-degenerate target state, say $|\Psi_{\rm A}\rangle$, and a trial state with multiple degenerate copies, say $\{|\Psi_{{\rm B},n}\rangle\}$, where $n$ is an index distinguishing between different degenerate copies.
One reasonable definition would be the projected amplitude of $|\Psi_{\rm A}\rangle$ onto the degenerate Hilbert space expanded by $\{|\Psi_{{\rm B},n}\rangle\}$.
Mathematically, the square of overlap defined in this way can be written as $|{\cal O}|^2= \sum_n |\langle \Psi_{\rm A}|\Psi_{{\rm B},n}\rangle |^2$.
We use precisely this definition here and in the following subsection, where the overlap between $\Psi_{7/3}[\delta V_1^{(1)}]$ and ${\cal C}_{\rm PH} \Psi_{Z_4}$ is discussed.

In the case of the fermionic Haffnian state, there is an additional serious problem that the ground-state degeneracy diverges as a function of particle number.
In this work, we focus only on supposedly the sub-Hilbert space of $\Psi_{\rm Hf}$, which can be represented by the antisymmetrized projections of the two degenerate Halperin (551) states at ${\bf Q}=(N/2,N/2)$, where $\Psi_{7/3}[\delta V_1^{(1)}]$ occurs.
In this way, we might neglect many other degenerate Haffnian states, which cannot be obtained by antisymmetrizing the Halperin (551) state.
Thus, potentially, our results in the torus geometry could seriously underestimate the true overlap between $\Psi_{7/3}[\delta V_1^{(1)}]$ and $\Psi_{\rm Hf}$.

To circumvent this problem, we have also performed ED in the spherical geometry, where the ground-state degeneracy issue does not occur. 
Postponing the detailed discussions to Sec.~\ref{sec:spherical_geometry}, it is sufficient for now to mention that, fortunately, the overlap results obtained in the spherical geometry are overall consistent with those in the torus geometry reported below.

Figure~\ref{fig:overlap} shows the overlap between $\Psi_{7/3}[\delta V_1^{(1)}]$ and $\Psi_{\rm Hf}$ in the torus geometry, which is sizable at moderately negative $\delta V_1^{(1)}/V_1^{(1)}$ for small particle numbers, say, $N=6$ and 8.
Unfortunately, the overlap decreases fast as the particle number increases to $N=10$. 
In particular, the overlap becomes negligibly small near the Coulomb point at $N=10$. 
We therefore conclude that $\Psi_{\rm Hf}$ has little chance of representing the exact 7/3 ground state around the Coulomb point in the thermodynamic limit.

Third, the overlap between $\Psi_{7/3}[\delta V_1^{(1)}]$ and ${\cal A} \Psi_{^6{\rm CFS}\otimes ^6{\rm CFS}}$ is somewhat similar to that between $\Psi_{7/3}[\delta V_1^{(1)}]$ and $\Psi_{\rm Hf}$.
That is, the overlap between $\Psi_{7/3}[\delta V_1^{(1)}]$ and ${\cal A} \Psi_{^6{\rm CFS}\otimes ^6{\rm CFS}}$ is sizable at moderately negative $\delta V_1^{(1)}/V_1^{(1)}$ at $N=8$, but collapses almost completely at $N=12$ for the entire range of $\delta V_1^{(1)}/V_1^{(1)}$.
Therefore, we also conclude that ${\cal A} \Psi_{^6{\rm CFS}\otimes ^6{\rm CFS}}$ has little chance of representing the exact 7/3 ground state around the Coulomb point in the thermodynamic limit.

Finally, let us discuss the overlap between $\Psi_{7/3}[\delta V_1^{(1)}]$ and ${\cal C}_{\rm PH} \Psi_{Z_4}$.
It has been mentioned in Sec.~\ref{sec:C_PH_Z_4} that there is the double ground-state degeneracy for $\Psi_{Z_4}$ at ${\bf Q} = (0,0)$.
After PH conjugation, ${\cal C}_{\rm PH} \Psi_{Z_4}$ has the double ground-state degeneracy at ${\bf Q}=(N/2, N/2)$ and $(0,0)$ for even and odd particle numbers, respectively, which can be taken as the {\it zero physical momentum}~\cite{Haldane85}.
As explained above, we define the overlap in this situation as the projected amplitude of $\Psi_{7/3}[\delta V_1^{(1)}]$ onto the degenerate Hilbert space expanded by the two degenerate copies of  ${\cal C}_{\rm PH} \Psi_{Z_4}$.

Figure~\ref{fig:overlap} shows that, while fluctuating somewhat across various particle numbers, the overlap between $\Psi_{7/3}[\delta V_1^{(1)}]$ and ${\cal C}_{\rm PH} \Psi_{Z_4}$ remains substantial around the Coulomb point.
It is interesting to note that the overlap between $\Psi_{7/3}[\delta V_1^{(1)}]$ and ${\cal C}_{\rm PH} \Psi_{Z_4}$ seems to be peaked around the Coulomb point, exactly where the Laughlin state loses its overlap significantly.  
Therefore, we conclude that ${\cal C}_{\rm PH} \Psi_{Z_4}$ has a reasonably good chance of representing the exact 7/3 ground state around the Coulomb point with its overlap being the highest among the four trial states.

\subsubsection{Spherical geometry}
\label{sec:spherical_geometry}

Now, we discuss what happens in the spherical geometry.
In particular, we are interested in the overlap between the exact 7/3 ground state at the Coulomb point and each of the following three trial states, $\Psi_{333}$, $\Psi_{\rm Hf}$, and ${\cal C}_{\rm PH} \Psi_{Z_4}$.
We do not consider ${\cal A} \Psi_{^6{\rm CFS}\otimes ^6{\rm CFS}}$ since its overlap with the exact 7/3 ground state is already too low in the torus geometry.
We do not think that this situation would change in the spherical geometry.

Before computing the overlap with the exact 7/3 ground state, we first check if $\Psi_{\rm Hf}$ is a worthy trial state. 
Specifically, we compute the Coulomb interaction energies of $\Psi_{333}$ and $\Psi_{\rm Hf}$ by performing Monte Carlo simulation up to $N=50$, the results of which are then extrapolated to the thermodynamic limit. 
To make such large-scale Monte Carlo simulation possible, it is important to write the trial states in concrete mathematical forms.
Unfortunately, the mathematical forms of $\Psi_{333}$ [Eq.~\eqref{eq:333}] and $\Psi_{\rm Hf}$ [Eq.~\eqref{eq:Haffnian}] are concretely known only in the LLL.
To perform Monte Carlo simulation in the SLL, we follow the trick invented in Ref.~\cite{Park98}. 
In this trick, the Coulomb potential in the SLL is modeled as the effective potential $V_{\rm eff}(r)=\frac{1}{r}+a_1e^{-\alpha_1 r^2}+a_2r^2e^{-\alpha_2 r^2}$, where $a_1$, $a_2$, $\alpha_1$, and $\alpha_2$ are fixed by requiring that the first four pseudopotentials of $V_{\rm eff}(r)$ in the LLL should be equal to those of the Coulomb potential in the SLL. 
As a result of this trick, we are able to estimate that, in the thermodynamic limit, the Coulomb interaction energies of $\Psi_{333}$ and $\Psi_{\rm Hf}$ are $-0.325(0)$ and $-0.320(9)$ in units of $e^2/\epsilon l_B$, respectively.
This means that the two trial states are very competitive in the SLL.
For comparison, note that, in the LLL, the Coulomb interaction energies of $\Psi_{333}$ and $\Psi_{\rm Hf}$ are estimated to be $-0.4097(3)$ and $-0.3719(1)$ in units of $e^2/\epsilon l_B$, respectively.

Now, we discuss the overlap between the exact 7/3 ground state at the Coulomb point and each of the three trial states, $\Psi_{333}$, $\Psi_{\rm Hf}$, and ${\cal C}_{\rm PH} \Psi_{Z_4}$, in the spherical geometry. 
To this end, it is important to mention that the three trial states are actually compared with their respective, slightly different exact 7/3 ground states due to the well-known property of the spherical geometry known as the ``flux shift.''
In the spherical geometry, monopole strength $Q$ is related with particle number $N$ for a desired thermodynamic filling factor $\nu$ via $2Q=\nu^{-1} N -S$. 
A problem is that different values should be assigned to flux shift $S$ for different trial states; $S=3$ for $\Psi_{333}$, $5$ for $\Psi_{\rm Hf}$, and $-3$ for ${\cal C}_{\rm PH} \Psi_{Z_4}$.
We define the exact 7/3 ground state as the exact Coulomb ground state occurring at appropriate $S$, which depends on the specific trial state.

It is interesting to mention that, when $S$ is set equal to 5 (which corresponds to $\Psi_{\rm Hf}$), the ground state energy exhibits a clear even-odd effect meaning that the ground state energy oscillates depending on whether the particle number is even or odd~\cite{Lu10}. 
This suggests that a certain form of the pairing correlation exists in $\Psi_{\rm Hf}$. 
When $S$ is set equal to 3 (which corresponds to $\Psi_{333}$), no such even-odd effect is observed.

\begin{figure}[t]
\includegraphics[width=0.43\textwidth]{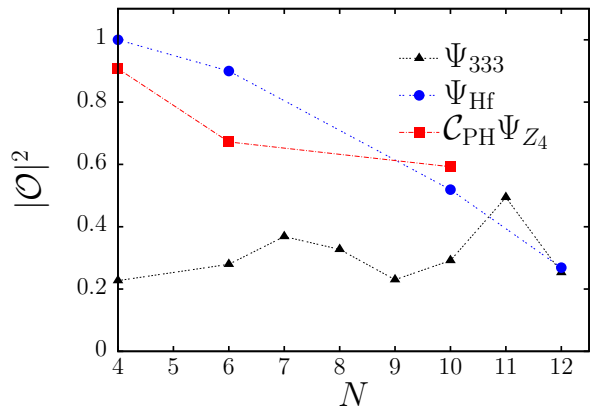}
\caption{(Color online) 
Square of overlap $|{\cal O}|^2$ between the exact 7/3 ground state at the Coulomb point and each of the three trial states, $\Psi_{333}$, $\Psi_{\rm Hf}$, and ${\cal C}_{\rm PH} \Psi_{Z_4}$, in the spherical geometry. 
The three trial states occur at different values of flux shift; $S=3$ for $\Psi_{333}$, $5$ for $\Psi_{\rm Hf}$, and $-3$ for ${\cal C}_{\rm PH} \Psi_{Z_4}$. 
Some data points are missing due to the fact that the total angular momentum of the ground state is non-zero there.
}
\label{fig:overlap_sphere}
\end{figure}

Figure~\ref{fig:overlap_sphere} shows the square of overlap between the exact 7/3 ground state at the Coulomb point and each of the three trial states, $\Psi_{333}$, $\Psi_{\rm Hf}$, and ${\cal C}_{\rm PH} \Psi_{Z_4}$, in the spherical geometry.
First, the overlap between the exact 7/3 ground state and $\Psi_{333}$ is fairly low, $20\mbox{-}50\%$, exhibiting a large undulation as a function of particle number.
Second, the overlap between the exact 7/3 ground state and $\Psi_{\rm Hf}$ is initially quite high, but decreases fast as the particle number increases.
This behavior is consistent with what is observed in the torus geometry.
Therefore, we arrive at the same conclusion as obtained in the torus geometry that both $\Psi_{333}$ and $\Psi_{\rm Hf}$ would have little chance of representing the exact 7/3 ground state around the Coulomb point in the thermodynamic limit.

Meanwhile, the overlap between the exact 7/3 ground state and ${\cal C}_{\rm PH} \Psi_{Z_4}$ is reasonably high, $60\mbox{-}90\%$, with a relatively stable behavior as a function of particle number.
Thus, it is concluded that, at least among the three trial states studied above, ${\cal C}_{\rm PH} \Psi_{Z_4}$ is the best candidate state for the exact 7/3 ground state around the Coulomb point.

%%%%%%%%%%%%%%%%%%%%%%%%%%%%%%%%%%%%%%%%%%%%%
%%%%%%%%%%%%%%%%%%%%%%%%%%%%%%%%%%%%%%%%%%%%%

%%%%%%%%%%%%%%%%
\subsection{Energy spectrum}
\label{sec:Energy_spectrum}
%%%%%%%%%%%%%%%%

The reasonably substantial overlap between the exact 7/3 ground state and ${\cal C}_{\mathrm{PH}}\Psi_{Z_4}$ around the Coulomb point motivates us to investigate if there is a signature in the exact energy spectrum supporting ${\cal C}_{\mathrm{PH}}\Psi_{Z_4}$.
A particular signature that we would like to focus on is the characteristic ground-state degeneracy of ${\cal C}_{\mathrm{PH}}\Psi_{Z_4}$ occurring at specific pseudomomentum channels in the torus geometry.

The ground-state degeneracy of $\Psi_{Z_4}$ can be evaluated by using two different methods.
The first method is to examine the structure of the conformal field theory~\cite{Read99}.
The second method is to examine the root configurations of $\Psi_{Z_4}$ in the thin-torus limit~\cite{Ardonne08,Liu15}, as discussed in Sec.~\ref{sec:C_PH_Z_4}.
The ground-state degeneracy of ${\cal C}_{\mathrm{PH}}\Psi_{Z_4}$ can be inferred from that of $\Psi_{Z_4}$ with the knowledge of how the root configurations transform under the PH conjugation.

By carefully examining the pseudomomenta allowed for all the root configurations of ${\cal C}_{\mathrm{PH}}\Psi_{Z_4}$, one can show that, for even particle numbers, there should be five degenerate ground states with one occurring at each ${\bf Q}=(0,0)$, $(N/2, 0)$, and $(0, N/2)$, and two at $(N/2, N/2)$.
This makes the total ground-state degeneracy fifteen since each of the above five degenerate ground states has three center-of-mass-shifted copies.
As a consequence, if ${\cal C}_{\rm PH} \Psi_{Z_4}$ were to represent the exact 7/3 ground state accurately, there should be a set of five degenerate, or at least quasidegenerate copies of the ground state with one occurring at each ${\bf Q}=(0,0)$, $(N/2, 0)$, $(0, N/2)$ and two at $(N/2, N/2)$, accompanied by their center-of-mass-shifted copies.

%%%%%%%%%%%%%%%%%%%%%%%%%%%%%%%%%%%%%%%%%%%%%%%%%%%%%%%%%%%%%%%%%
\begin{figure}[t]
\includegraphics[width=0.4\textwidth]{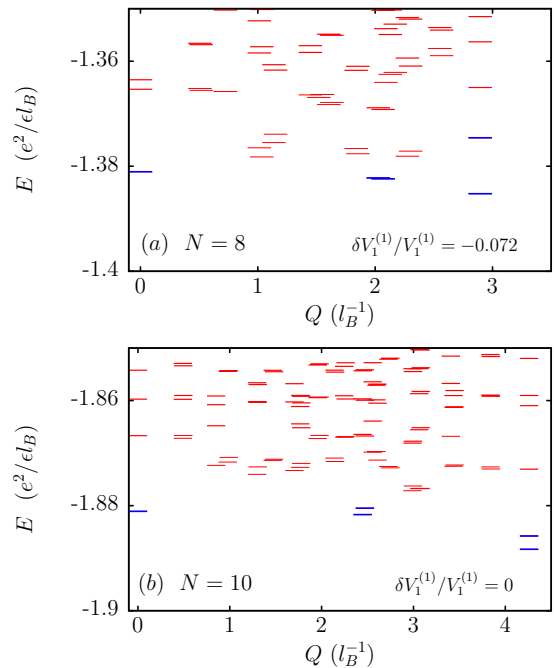}
\caption{
(Color online) 
Exact energy spectra in the torus geometry as a function of the magnitude of pseudomomentum $Q=|{\bf Q}|$ in units of $1/l_B$.
Note that the unit cells are rectangular and hexagonal in Panels (a) and (b), respectively.
The lowest-energy states at pseudomomenta ${\bf Q}=(0,0)$, $(N/2, 0)$, $(0, N/2)$, and $(N/2, N/2)$ are denoted as blue lines, distinguished from all the other states denoted as red lines.
Note that, here, slightly non-unity aspect ratios are used to separate the two, otherwise degenerate states at ${\bf Q}=(N/2,0)$ and $(0,N/2)$ denoted as the two blue lines in the middle.
}
\label{fig:degeneracy}
\end{figure}
%%%%%%%%%%%%%%%%%%%%%%%%%%%%%%%%%%%%%%%%%%%%%%%%%%%%%%%%%%%%%%%%%

Figure~\ref{fig:degeneracy} shows the exact energy spectra in the torus geometry as a function of the magnitude of pseudomomentum $Q=|{\bf Q}|$ in units of $1/l_B$.
Considering that the anticipated quasi-degeneracy could be more visible when the overlap between the exact 7/3 ground state and ${\cal C}_{\rm PH} \Psi_{Z_4}$ is relatively high, we choose the $N=8$ system in Panel (a) with $\delta V^{(1)}_1/V^{(1)}_1= -0.072$ and the rectangular unit cell, corresponding to the regime of maximum overlap in Panel (b) in Fig.~\ref{fig:overlap}.
A slightly non-unity aspect ratio, $r_a=0.98$, is chosen so that the two, otherwise degenerate states at ${\bf Q}=(N/2,0)$ and $(0,N/2)$ are separated. 
Similarly, we choose the $N=10$ system in Panel (b) with $\delta V^{(1)}_1/V^{(1)}_1 = 0$ and the hexagonal unit cell, corresponding to the regime of maximum overlap in Panel (g) in Fig.~\ref{fig:overlap}. 
Here, the aspect ratio is chosen to be $r_a=0.99$.
As one can see from the figure, there is a reasonably strong signature for the anticipated quasi-degeneracy of the ground state at the correct pseudomomentum channels.

%%%%%%%%%%%%%%%%%%%%%%%%%%%%%%%%%%%%%%%%%%%%%%%%%%%%%%%%%%%%%%%%%
\begin{figure}[t]
\includegraphics[width=0.48\textwidth]{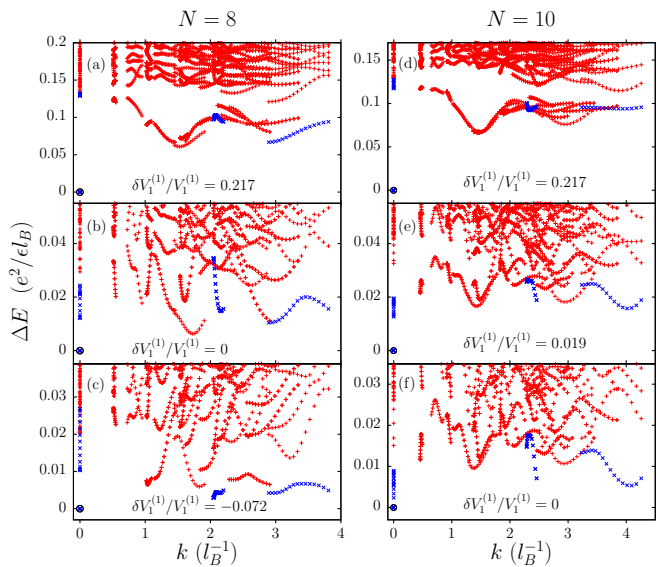}
\caption{
(Color online) Exact energy spectra in the torus geometry for various values of the Haldane pseudopotential variation $\delta V^{(1)}_1/V^{(1)}_1$ as a function of the magnitude of physical momentum $k=|{\bf k}|$ in units of $1/l_B$. 
The physical momentum ${\bf k}$ is related with the pseudomomentum ${\bf Q}$ via ${\bf k}= {\bf Q} - {\bf k}_0$, where ${\bf k}_0$ denotes the zero momentum.
The lowest-energy states at pseudomomenta ${\bf Q}=(0,0)$, $(N/2, 0)$, $(0, N/2)$, and $(N/2, N/2)$ are denoted as blue x's in comparison with all the other states denoted as red crosses.
Note that, here, the aspect ratio is chosen to be unity so that all the energy eigenstates at ${\bf Q}=(N/2,0)$ and $(0,N/2)$ are exactly superposed.
}
\label{fig:energy_spectrum}
\end{figure}
%%%%%%%%%%%%%%%%%%%%%%%%%%%%%%%%%%%%%%%%%%%%%%%%%%%%%%%%%%%%%%%%%

Now, we examine the energy spectrum from a more physical point of view.
Specifically, we investigate the change in the energy spectrum by moving from sufficiently large positive $\delta V^{(1)}_1/V^{(1)}_1$, where the Laughlin state is relevant, to the Coulomb point, where the PH-conjugated $Z_4$ parafermion state is anticipated to be relevant.

Figure~\ref{fig:energy_spectrum} shows the exact energy spectra in the torus geometry for various values of the Haldane pseudopotential variation $\delta V^{(1)}_1/V^{(1)}_1$ as a function of the magnitude of physical momentum $k=|{\bf k}|$ in units of $1/l_B$.
The physical momentum ${\bf k}$ is related with ${\bf Q}$ via ${\bf k}= {\bf Q} - {\bf k}_0$, where ${\bf k}_0$ denotes the zero momentum.
In our situation, ${\bf k}_0=( N^\prime/2, N^\prime/2 )$ and $(0,0)$ for even and odd $N^\prime$, respectively, where $N^\prime=\mathrm{gcd}(N,N_{\phi})$~\cite{Haldane85}.
This means that ${\bf k}_0=(N/2, N/2)$ in Fig.~\ref{fig:energy_spectrum} since $N^\prime=N$ is even. 
The energy spectra are computed in two different finite-size systems with $N=8$ [Panel (a), (b), and (c) ] and 10 [Panel (d), (e), and (f)], both of which show overall similar behaviors as a function of $\delta V^{(1)}_1/V^{(1)}_1$.

At sufficiently large positive $\delta V^{(1)}_1/V^{(1)}_1$, the energy spectrum exhibits a well-developed magnetoroton structure with its minimum located at $|{\bf k}|/l_B \simeq 1.4$, which is the defining signature of the Laughlin state~\cite{Haldane83, Haldane85, Haldane85_Finite-size, Girvin85}. 
This is consistent with the fact that the overlap between $\Psi_{7/3}[\delta V^{(1)}_1]$ and $\Psi_{333}$ is essentially unity at sufficiently large positive $\delta V^{(1)}_1/V^{(1)}_1$.

On the other hand, as $\delta V^{(1)}_1/V^{(1)}_1$ is lowered to the Coulomb point, the energy spectrum undergoes an intriguing transition from the spectrum with the Laughlin-type magnetoroton structure to that with the specific quasi-degeneracy of the ground state, which is characteristic to the PH-conjugated $Z_4$ parafermion state, as discussed above. 
That is, the lowest-energy states at ${\bf Q}=(0,0)$, $(N/2, 0)$, $(0, N/2)$, and $(N/2, N/2)$ are pulled away from the continuum of other excited states to become essentially the lowest-energy excited states at the Coulomb point or slightly negative  $\delta V_1^{(1)}$.
We believe that this provides reasonably strong evidence supporting that the PH-conjugated $Z_4$ parafermion state provides a good trial state representing the exact 7/3 ground state around the Coulomb point.

It is worthwhile to mention that the quasidegenerate excited state at ${\bf k}=0$ was previously interpreted as the onset of the incompressible-to-compressible phase transition~\cite{Haldane_Book}.
In our interpretation, this state is one of the two degenerate copies of the $Z_4$ parafermion state occurring at ${\bf Q}=(N/2,N/2)$. 
%along with the others at ${\bf Q}=(0,0)$, $(N/2, 0)$, and $(0, N/2)$.

\section{Conclusion}
\label{sec:Conclusion}

In this work, we investigate the nature of the FQH state at $\nu=7/3$ (and $8/3$ in the presence of the PH symmetry) by using ED in both torus and spherical geometries.  
Specifically, we compute the overlap between the exact $7/3$ ground state and various competing states including (i) the Laughlin state, $\Psi_{333}$, (ii) the fermionic Haffnian state, $\Psi_{\rm Hf}$, which is proven in this work to be entirely equivalent to ${\cal A}\Psi_{551}$, (iii) the antisymmetrized product state of two CF seas at 1/6 filling, ${\cal A} \Psi_{^6{\rm CFS}\otimes ^6{\rm CFS}}$, and (iv) the PH-conjugated $Z_4$ parafermion state, ${\cal C}_{\rm PH} \Psi_{Z_4}$, with $\Psi_{Z_4}$ identified as ${\cal A} \Psi_{330}$ under the proper understanding of the ground-state degeneracy in the torus geometry.

It is shown that, while valid at sufficiently large positive Haldane pseudopotential variation $\delta V_1^{(1)}$, the Laughlin state loses its overlap with the exact 7/3 ground state significantly around the Coulomb point.
At slightly negative  $\delta V_1^{(1)}$, the PH-conjugated $Z_4$ parafermion state is shown to have a substantial overlap with the exact 7/3 ground state, being the highest among the above four trial states. 
Also, around the Coulomb point, the energy spectrum exhibits an intriguing change from the spectrum with the Laughlin-type magnetoroton structure to that with the specific quasi-degeneracy of the ground state, which is characteristic to the PH-conjugated $Z_4$ parafermion state. 
Therefore, we conclude that the PH-conjugated $Z_4$ parafermion state has a reasonably good chance of representing the exact 7/3 ground state around the Coulomb point.

From the perspective of the general guiding principle for the FQH states in the SLL, all the above trial states are constructed according to a guiding principle called the bilayer mapping approach, where a trial state is obtained as the antisymmetrized projection of a bilayer quantum Hall state with interlayer distance $d$ as a variational parameter. 
The bilayer mapping approach can be regarded as an alternative to the $Z_k$ parafermion approach, while the two approaches coincide in the case of the $Z_4$ parafermion state.
That is to say, $\Psi_{Z_4} = {\cal A} \Psi_{330}$.

It is emphasized that, in the torus geometry, there is an issue of the ground-state degeneracy mismatch between $\Psi_{Z_4}$ and ${\cal A} \Psi_{330}$.
The ground-state degeneracy mismatch is in apparent conflict with the analytical proof that the wave functions of the two states are exactly identical in the infinite planar geometry~\cite{Rezayi10,Barkeshli10}. 
This apparent conflict is resolved by interpreting the analytical proof in such a way that, in the torus geometry, $\Psi_{Z_4}$ and ${\cal A} \Psi_{330}$ are exactly identical only in the common momentum sector, where both have the same ground-state degeneracy.
We have explicitly confirmed this to be the case by showing that the entire degenerate Hilbert space expanded by the degenerate copies of $\Psi_{Z_4}$ is exactly identical to that of ${\cal A} \Psi_{330}$ at ${\bf Q}=(0,0)$ [mod $\textrm{gcd}(N, N_\phi)$]. %where both states have the same ground-state degeneracy.

An interesting future direction is to investigate if the bilayer mapping approach can be applied to the 2/5-filled SLL. 
Several trial states are natural in terms of the bilayer mapping approach.
%which are the antisymmetrized projections of the Halperin (441) and (550) states, ${\cal A} \Psi_{441}$ and ${\cal A} \Psi_{550}$, at $\nu=12/5$ and their PH conjugates at $\nu=13/5$.
Particularly, in light of the substantial overlap between the exact 7/3 ground state and ${\cal C}_{\rm PH} \Psi_{Z_4} (={\cal C}_{\rm PH} {\cal A} \Psi_{330})$, it would be interesting to compute the overlap between the exact 12/5 ground state and the antisymmetrized Halperin (550) state, ${\cal A} \Psi_{550}$.
Note that the antisymmetrized Halperin $(nn0)$ state, ${\cal A} \Psi_{nn0}$, was considered previously~\cite{Rezayi10,Barkeshli11}, while its overlap with the exact Coulomb ground state was not studied.

It was argued in previous numerical studies~\cite{Read99,Zhu15,Mong15} that the $Z_3$ parafermion state provides a good trial state representing the exact 13/5 ground state around the Coulomb point. 
This means that, in the absence of PH breaking, the PH-conjugated $Z_3$ parafermion state could provide a good trial state at $\nu=12/5$. 
On the other hand, it was shown previously~\cite{Bonderson12} that the Bonderson-Slingerland state is in a tight energy competition with the $Z_3$ parafermion state at $\nu=12/5$.
If so, it would be interesting to investigate which state among the antisymmetrized Halperin (550) state, the PH-conjugated $Z_3$ parafermion state, and the Bonderson-Slingerland state has the best overlap with the exact 12/5 ground state around the Coulomb point.
Note that the antisymmetrized Halperin (550) state by itself is expected to be gapless according to the pattern-of-zeros and vertex algebra approaches~\cite{Barkeshli11}, while the $Z_3$ parafermion state and the Bonderson-Slingerland state are known to be gapped.
The antisymmetrized Halperin (550) state, however, can be gapped with addition of perturbations driving the state away from the critical point~\cite{Barkeshli11}.

%%%%%%%%%%%%%%%%%%%%%%
\acknowledgements
%%%%%%%%%%%%%%%%%%%%%%

The authors are grateful to Jainendra K. Jain, Edward H. Rezayi, Steven H. Simon, Zlatko Papi\'{c}, Gun Sang Jeon, Gil Young Cho, and Sutirtha Mukherjee for their insightful discussions and useful comments. 
This work was supported by IBS-R009-D1 (Y1) and Supercomputing Center of Korea Institute of Science and Technology Information (KISTI) with supercomputing resources including technical support (KSC-2014-C3-033).
Also, we thank KIAS Center for Advanced Computation for providing computing resources.

%%%%%%%%%%%%%%%%%%%%%%%%%%%%%%%%%%%%%%%%%%%%%%%%%%%%%%%%%%
\appendix
%%%%%%%%%%%%%%%%%%%%%%%%%%%%%%%%%%%%%%%%%%%%%%%%%%%%%%%%%%

%%%%%%%%%%%%%%%%%%%%%%%%%%%%%%%%%%%%%%%%%%%%%%%%%%%%%%%%%%
\section{Fermionic $(k+1)$-body $\delta$-function interaction}
\label{appendix:k+1-body}

In this Appendix, we prove that the $Z_k$ parafermion state is the zero-energy ground state of the fermionic $(k+1)$-body $\delta$-function interaction at filling factor $\nu=k/(k+2)$, which is  written as
\begin{widetext}
\begin{align}
\label{eq:H_k+1}
H_{k+1} 
= \sum_{p_1<p_2<\cdots<p_{k+1}} {\cal S}_{p_1,p_2,\dots,p_{k+1}}
\left\{\nabla_{p_1}^{2k}\nabla_{p_2}^{2(k-1)}\cdots \nabla_{p_k}^{2}\right\} 
\delta^2({z}_{p_1}-{z}_{p_2}) \delta^2({z}_{p_2}-{z}_{p_3}) \cdots \delta^2({z}_{p_k}-{z}_{p_{k+1}}),
\end{align} 
\end{widetext}
where ${\cal S}_{p_1,p_2,\dots,p_{k+1}}$ is the symmetrization operator, and $\nabla^{2(k-j+1)}_{p_j}=\left(\partial/\partial z_{p_j}\right)^{k-j+1}\left(\partial/\partial \bar{z}_{p_j}\right)^{k-j+1}$.
Here, the $Z_k$ parafermion wave function is given by $\Psi_{Z_k} = J \psi_k$, where $J=\prod_{i<j}^N (z_i-z_j)$ and $\psi_k$ is the bosonic Read-Rezayi (RR) wave function, which is defined as the zero-energy ground state of the bosonic $(k+1)$-body $\delta$-function interaction at $\nu=k/2$~\cite{Read99}, 
\begin{widetext}
\begin{align}
\label{eq:bosonic_H_k+1}
H_{k+1}^{\rm boson} = \sum_{p_1<p_2<\cdots<p_{k+1}} \delta^2({z}_{p_1}-{z}_{p_2}) \delta^2({z}_{p_2}-{z}_{p_3}) \cdots \delta^2({z}_{p_k}-{z}_{p_{k+1}}).
\end{align}
\end{widetext}

To begin with, we compute the energy expectation value of $H_{k+1}$ for a trial fermionic wave function $\Psi=J\psi$ at $\nu=k/(k+2)$ with $\psi$ being an arbitrary bosonic wave function at $\nu=k/2$.
Then, we show that the energy expectation value becomes zero if and only if $\psi=\psi_k$, i.e., the bosonic RR wave function.
Since $H_{k+1}$ cannot have a negative expectation value, $\Psi_{Z_k}=J\psi_k$ is not only the zero-energy state, but also the ground state of $H_{k+1}$. 
In the infinite planar geometry, the uniqueness of $\Psi_{Z_k}$ at $\nu=k/(k+2)$ is guaranteed by that of $\psi_k$ at $\nu=k/2$~\cite{Read99}.
It is worthwhile to mention that, written in the form of a Trugman-Kivelson-type interaction, the above expression for the fermionic $(k+1)$-body $\delta$-function interaction was previously mentioned in Ref.~\cite{Lee15}, while the power of each differential operator was not specified explicitly. %in contrast to Eq.~\eqref{eq:H_k+1}.

For the sake of convenience, let us rewrite Eq.~\eqref{eq:H_k+1} in a compact form:
\begin{align}
\label{eq:H_k+1_compact}
H_{k+1}& = \sum_{p\in S(N, k+1)} {\cal S}_p \prod_{j=1}^{k} \nabla_{p_j}^{2j} \delta^2_p(\{z\}), 
\end{align}
where $S(m,n)$ indicates all possible sets of $n$ different indices chosen from $\left\{1,2,\dots, m\right\}$, 
$N$ is the total number of particles,
${\cal S}_p \equiv {\cal S}_{p_1,p_2,\dots,p_{k+1}}$, 
$\nabla^{2j}_{p_j} = (\partial / \partial z_{p_j})^j (\partial/ \partial \bar{z}_{p_j})^j \equiv \partial^{j}_{p_j} \bar{\partial}^{j}_{p_j}$,
and $\delta^2_p(\{z\}) \equiv \delta^2({z}_{p_1}-{z}_{p_2})\delta^2({z}_{p_2}-{z}_{p_3})\cdots \delta^2({z}_{p_k}-{z}_{p_{k+1}})$. 
Note that the powers of differential operators can be rearranged as shown in Eq.~\eqref{eq:H_k+1_compact} due to the symmetrization operator.

The energy expectation value of $H_{k+1}$ for $\Psi=J\psi$ is written as follows: 
\begin{align}
& \langle \Psi |H_{k+1} |\Psi \rangle \nonumber \\
& =\sum_{p\in S(N, k+1)} \int  {\cal S}_p \left(\prod_{j=1}^{k} \nabla_{p_j}^{2j}\delta^2_p(\{z\}) \right) \bar{\Psi} \Psi  \label{eq:energy[2]} \\
& = (k+1)! \sum_{p\in S(N,k+1)} \int  \left( \prod_{j=1}^{k} \nabla_{p_j}^{2j}\delta^2_p(\{z\}) \right) \bar{\Psi} \Psi     \label{eq:energy[3]} \\
& = (k+1)! \sum_{p\in S(N,k+1)} \int  \delta^2_p(\{z\})  \left(\prod_{j=1}^{k} \nabla_{p_j}^{2j} \bar{\Psi} \Psi \right) ,  \label{eq:energy[4]} 
\end{align}
where the differential $\prod_{i}dx_idy_i= {2^{-N}}\prod_{i}d\bar{z}_idz_i$ is omitted for simplicity. 
Since the spatial coordinates are dummy variables, which are to be integrated out, the symmetrization operator just generates $(k+1)!$ identical terms as shown in Eq.~\eqref{eq:energy[3]}. 
Then, performing integration by parts, one can obtain Eq.~\eqref{eq:energy[4]}, where the term inside the parenthesis is further evaluated by using the fact that $\Psi = J\psi$:
\begin{align}
& \prod_{j=1}^{k}\nabla_{p_j}^{2j}\bar{\Psi}\Psi \nonumber \\
&= \prod_{j=1}^{k} \partial_{p_j}^{j}\bar{\partial}_{p_j}^{j} \left(\bar{J}\bar{\psi} J\psi\right) \label{eq:differential_operator[2]} \\
& = \left(\prod_{j=1}^{k}\bar{\partial}^j_{p_j}\bar{J_p}\right)\left(\prod_{j=1}^{k}\partial^j_{p_j}J_p\right)\left|{J\over J_p}\right|^2 \left|\psi \right|^2+ \chi_p \label{eq:differential_operator[3]} \\
& = \prod_{j=1}^{k} (j!)^2 \left|{J\over J_p}\right|^2 \left|\psi \right|^2+ \chi_p, \label{eq:differential_operator[4]}
\end{align}
where $J_p = \prod_{1 \le i < j \le k+1}(z_{p_i} - z_{p_j})$ and 
$\chi_p$ denotes any terms generated by applying $\partial_{p_j}$ and $\bar{\partial}_{p_j}$ at least once to $\psi$ rather than $J_p$ and $\bar{J}_p$, respectively.

Now, the $\delta$-function constraint in Eq.~\eqref{eq:energy[4]} imposes $z_{p_1}=z_{p_2} = \cdots = z_{p_{k+1}}$, which makes $\chi_p$ vanish since at least one factor from $J_p$ survives in the form of $(z_{p_i}-z_{p_j})$.
On the other hand, the first term in Eq.~\eqref{eq:differential_operator[4]} does not necessarily vanish, giving rise to a non-negative energy expectation value in Eq~\eqref{eq:energy[4]}:
\begin{align} 
\langle \Psi |H_{k+1} |\Psi \rangle  \propto \sum_{p\in (N,k+1)} \int  \delta^2_p(\{z\}) \left|\psi \right|^2\left| \frac{J}{J_p} \right|^2.
\end{align} 
The energy expectation value becomes zero if and only if $\psi$ vanishes when $z_{p_1}=z_{p_2}=\cdots=z_{p_{k+1}}$. 
This means that $\psi$ is none other than the bosonic RR wave function, $\psi_k$. 
Consequently, the $Z_k$ parafermion wave function, $\Psi_{Z_k}= J \psi_k$, is the zero-energy ground state of the fermionic $(k+1)$-body $\delta$-function interaction.
%In the infinite planar geometry, the uniqueness of $\Psi_{Z_k}$ at $\nu=k/(k+2)$ is guaranteed by that of $\psi_k$ at $\nu=k/2$.
Q.E.D.

%%%%%%%%%%%%%%%%%%%%%%%%%%%%%%%%%%%%%%%%%%%%%%%%%%%%%%%%%%
\section{Second quantization form of the fermionic five-body $\delta$-function interaction in the torus geometry}
\label{appendix:five-body}

Here, we describe how to obtain the second quantization form of the fermionic five-body $\delta$-function interaction in the torus geometry. 
%and its second-quantization representation in terms of the torus basis states.  
Note that this description can be generalized to any fermionic (bosonic) $(k+1)$-body $\delta$-function interaction.

To begin with, the fermionic five-body $\delta$-function interaction is written in the planar geometry as follows:
\begin{widetext}
\begin{align}
\label{eq:H_5_planar}
H_{\rm 5} & =\sum_{i<j<k<l<s}{\cal S}_{i,j,k,l,s}\left[\nabla^8_i\nabla^6_j\nabla^4_k \nabla^2_l\delta^2({\bf r}_i-{\bf r}_j)\delta^2({\bf r}_j-{\bf r}_k)\delta^2({\bf r}_k-{\bf r}_l)\delta^2({\bf r}_l-{\bf r}_s)\right] \\
&\equiv \sum_{i<j<k<l<s} {\cal S}_{i,j,k,l,s}V_{\rm 5}\left({\bf x}_1, {\bf x}_2, {\bf x}_3, {\bf x}_4\right),
\end{align}
\end{widetext}
where  ${\cal S}_{i,j,k,l,s}$ is the symmetrization operator, 
${\bf x}_1 = {\bf r}_i - {\bf r}_j$, ${\bf x}_2 = {\bf r}_j - {\bf r}_k$, 
${\bf x}_3 = {\bf r}_k - {\bf r}_l$, and ${\bf x}_4 = {\bf r}_l - {\bf r}_s$.
In turn, $V_5\left({\bf x}_1, {\bf x}_2, {\bf x}_3, {\bf x}_4\right)$ can be rewritten in terms of its Fourier components: 
\begin{widetext}
\begin{align}
\label{eq:V_5_planar}
V_{5} ( {\bf x}_1, {\bf x}_2, {\bf x}_3, {\bf x}_4 ) 
%& =  {1\over (2\pi)^8} 
%\int_{\rm 2D}d{\bf q}_1 \int_{\rm 2D} d{\bf q}_2 \int_{\rm 2D} d{\bf q}_3 \int_{\rm 2D} d{\bf q}_4
%|{\bf q}_1|^8 |{\bf q}_5|^6 |{\bf q}_6|^4 |{\bf q}_7|^2  
%\exp{\left(i\sum_{{\nu}=1}^4{\bf q}_{\nu}\cdot {\bf x}_{\nu} \right)}
%\nonumber \\
& = \int_{\rm 2D}d^2{\bf q}_1 \int_{\rm 2D} d^2{\bf q}_2 \int_{\rm 2D} d^2{\bf q}_3 \int_{\rm 2D} d^2{\bf q}_4
{\cal V}_5({\bf q}_1,{\bf q}_2,{\bf q}_3,{\bf q}_4)
\exp{\left(i\sum_{{\nu}=1}^4{\bf q}_{\nu}\cdot {\bf x}_{\nu} \right)} , 
\end{align}
\end{widetext}
where $\int_{\rm 2D}$ represents the integration over the infinite two-dimensional space,
and ${\cal V}_5({\bf q}_1,{\bf q}_2,{\bf q}_3,{\bf q}_4)=|{\bf q}_1|^8 |{\bf q}_5|^6 |{\bf q}_6|^4 |{\bf q}_7|^2 / (2\pi)^8$ 
with ${\bf q}_5 = {\bf q}_2 - {\bf q}_1$, ${\bf q}_6 = {\bf q}_3 - {\bf q}_2$, and ${\bf q}_7 = {\bf q}_4 - {\bf q}_3$.

The fermionic five-body $\delta$-function interaction in the torus geometry, $H_5^{\rm torus}$, is obtained by modifying $H_5$ to satisfy the periodic boundary condition with respect to the unit cell defined by two vectors, ${\bf L}_1 = L_1(\sin{\theta} \hat{x} +\cos{\theta} \hat{y})$ and ${\bf L}_2 = L_2 \hat{y}$:
\begin{align}
\label{eq:H_5_torus}
H_5^{\rm torus} = \sum_{i<j<k<l<s} {\cal S}_{i,j,k,l,s} V^{\rm torus}_5 ({\bf x}_1, {\bf x}_2, {\bf x}_3, {\bf x}_4), 
\end{align}
where 
%\begin{widetext}
\begin{align}
\label{eq:V_5_torus}
V_5^{\rm torus}(\{{\bf x}_\nu\}) =  \sum_{\nu=1}^{4}\sum_{u_\nu,v_\nu = -\infty}^{\infty}
V_{5}(\{ {\bf x}_\nu+u_\nu {\bf L}_1+v_\nu {\bf L}_2 \}),
\end{align}
%\end{widetext}
where $\{ {\bf x}_\nu \} = ( {\bf x}_1, {\bf x}_2, {\bf x}_3, {\bf x}_4 )$. %in Eq.~\eqref{eq:V_5_planar}.

The periodic boundary condition in Eq.~\eqref{eq:V_5_torus} can be readily satisfied by choosing the following Fourier representation:
\begin{align}
V_{5}^{\rm torus}(\{{\bf x}_\nu\})
=\sum_{\{{\bf q}_\nu\}} {\cal V}^{\rm torus}_5 (\{{\bf q}_\nu\}) 
\exp{\left( i\sum_{{\nu}=1}^{4} {\bf q}_{\nu}\cdot{\bf x}_{\nu}\right)}, 
\end{align}
where ${\bf q}_{\nu} = m_{\nu}{\bf Q}_1 + n_{\nu} {\bf Q}_2$ $(m_{\nu}, n_{\nu} \in {\mathbb Z})$ with ${\bf Q}_1$ and ${\bf Q}_2$ being the reciprocal vectors that satisfy ${\bf L}_1\cdot{\bf Q}_1 = 2\pi$ and ${\bf L}_2 \cdot  {\bf Q}_2 = 2\pi$, respectively.
Note that the summation is taken over all possible integer values of $(m_\nu, n_\nu)$.
Here, ${\cal V}^{\rm torus}_5 (\{{\bf q}_\nu\})$ is the same as ${\cal V}_5(\{ {\bf q}_\nu\})$ except the normalization factor; ${\cal V}^{\rm torus}_5 (\{{\bf q}_\nu\})=|{\bf q}_1|^8 |{\bf q}_5|^6 |{\bf q}_6|^4 |{\bf q}_7|^2 / v^4$ where $v$ is the area of the unit cell, i.e., ${v} = |{\bf L}_1 \times {\bf L}_2|$

Now, the fermionic five-body $\delta$-function interaction can be represented in terms of the torus basis states:
\begin{widetext}
\begin{align}
\label{eq:second_quantized_fivebody[1]}
H_5^{\rm torus}
&= \frac{1}{5!} \sum_{j_1, \dots, j_{10}}\left\langle j_1j_2j_3j_4j_5 \left| {\cal S}_{i,j,k,l,s}V_5^{\rm torus} (\{{\bf x}_\nu\}) \right| j_{10}j_9j_8j_7j_{6}\right\rangle c_{j_1}^{\dagger}c_{j_2}^{\dagger}c_{j_3}^{\dagger}c_{j_4}^{\dagger}c_{j_5}^{\dagger}c_{j_6}c_{j_7}c_{j_8}c_{j_9}c_{j_{10}} \\
\label{eq:second_quantized_fivebody[2]}
&\equiv \sum_{j_1, \cdots, j_{10}} {\cal A}_{j_1,\cdots,j_{10}}~c_{j_1}^{\dagger}c_{j_2}^{\dagger}c_{j_3}^{\dagger}c_{j_4}^{\dagger}c_{j_5}^{\dagger}c_{j_6}c_{j_7}c_{j_8}c_{j_9}c_{j_{10}},
\end{align}
\end{widetext}
where $\left\langle \cdots | {\cal S}_{i,j,k,l,s}V_5^{\rm torus} | \cdots \right\rangle = 5! \left\langle \cdots|V_5^{\rm torus} |\cdots \right\rangle $ since the symmetrization operator just generates $5!$ identical terms.
Above, ${\cal A}_{j_1,\cdots,j_{10}}$ is defined by
\begin{align}
\label{eq:A[10]}
{\cal A}_{j_1,\cdots,j_{10}}
%&={1\over v^4} \sum_{\{{\bf q}_\nu\}} q_1^8 q_5^6 q_6^4 q_7^2 {\cal B}_{j_1,\cdots,j_{10}}, \\
&= \sum_{\{{\bf q}_\nu\}} {\cal V}_5^{\rm torus}(\{{\bf q}_\nu\}) {\cal B}_{j_1,\cdots,j_{10}}(\{{\bf q}_\nu\}) ,
\end{align}
where %$\{{\bf q}_\nu\}=( {\bf q}_1, {\bf q}_2, {\bf q}_3, {\bf q}_4 )$, and
\begin{align}
\label{eq:B[10]}
&{\cal B}_{j_1,\cdots,j_{10}} (\{{\bf q}_\nu\}) 
\nonumber \\
&=  
\langle j_1| \exp{(i{\bf q}_1\cdot {\bf r}_i)} |j_{10}\rangle 
\langle j_2 | \exp{(i{\bf q}_5\cdot {\bf r}_j)} | j_9\rangle 
\nonumber \\
&\times
\langle j_3 | \exp{(i{\bf q}_6\cdot {\bf r}_k)} | j_8\rangle 
\langle j_4 | \exp{(i{\bf q}_7\cdot{\bf r}_l)} | j_7\rangle 
\nonumber \\
&\times
\langle j_5 | \exp{(-i{\bf q}_4\cdot {\bf r}_s)} |  j_6\rangle ,
\end{align}
where $\{{\bf q}_\nu\}=( {\bf q}_1, {\bf q}_2, {\bf q}_3, {\bf q}_4 )$, and ${\bf q}_5$, ${\bf q}_6$, and ${\bf q}_7$ are derived from $\{{\bf q}_\nu\}$ as shown below Eq.~\eqref{eq:V_5_planar}. 

%$\{{\bf q}_\nu\}=( {\bf q}_1, {\bf q}_2, {\bf q}_3, {\bf q}_4 )$, ${\bf q}_5 = {\bf q}_2 - {\bf q}_1$, ${\bf q}_6 = {\bf q}_3 - {\bf q}_2$, and ${\bf q}_7 = {\bf q}_4 - {\bf q}_3$.

By using the single-particle eigenstate of the lowest Landau level,  
%\begin{widetext}
\begin{align}
\label{eq:single-particle-state}
\langle {\bf r} | j \rangle 
=& \left(1\over  L_2 l_B \sqrt{\pi} \right)^{1 \over 2}\sum_{d=-\infty}^{\infty} 
\exp{\left[{ i(X_{j}+d L_1 \sin\theta ) y\over l_B^2} \right] }
\nonumber \\
&\times \exp{\left[-{\left(x + X_{j} + d L_1\sin\theta \right)^2\over 2l_B^2}\right]}
\nonumber \\
&\times \exp{\left[-{ i \cos\theta\over 2 l_B^2  \sin\theta}\left(X_{j}+d L_1\sin\theta\right)^2\right]},
\end{align}
%\end{widetext}
where $X_{j} = {2\pi l_B^2 j / L_2}$ and $L_1L_2\sin\theta = 2\pi l_B^2 N_{\phi}$,
one can eventually obtain the following expression for ${\cal B}_{j_1,\cdots,j_{10}} (\{{\bf q}_\nu\})$:
\begin{widetext}
\begin{align}
\label{eq:B[20]}
&{\cal B}_{j_1,\cdots,j_{10}} (\{{\bf q}_\nu\}) 
\nonumber \\
&=\delta'_{j_1-j_{10},n_1}  \delta'_{j_2-j_9,n_5}\delta'_{j_3-j_8,n_6} \delta'_{j_6-j_5,n_4} \delta'_{j_1+j_2+j_3+j_4+j_5, j_6+j_7+j_8+j_9+j_{10}} 
\exp{\left[ -\frac{l_B^2}{4} (|{\bf q}_1|^2+|{\bf q}_4|^2+|{\bf q}_5|^2+|{\bf q}_6|^2+|{\bf q}_7|^2) \right]} 
\nonumber \\
& \times \exp{\left[ {i \pi \over N_{\phi}} \{m_1( n_2 + 2j_9 -2j_1 )+m_2(n_1-2j_9-n_3+2j_3)
+m_3(n_4+2j_7-n_2-2j_3)+ m_4(n_3-2j_7+2j_5)\} \right]} ,
\end{align}
\end{widetext}
where it is important to note that $m_\nu$ and $n_\nu$ ($\nu=1, \cdots, 4$) are related to ${\bf q}_{\nu}$ via ${\bf q}_{\nu} = m_{\nu}{\bf Q}_1 + n_{\nu} {\bf Q}_2$.
Also, the primed Kronecker delta is defined as $\delta'_{s,t}=1$ if $s=t$ modulo $N_\phi$ and $0$ otherwise.

%%%%%%%%%%%%%%%%%%%%%%%%%%%%%%%%%%%%%%%%%%%%%%%%%%%%%%%%%%%%%%%
\section{An identity for the symmetrized Jastrow factor}
\label{appendix:identity}

In this Appendix, we prove Eq.~\eqref{eq:symmetric_polynomial_1}.
We begin by setting $N/2=n$, $x_i=z_i$, $x_{i+n}=\omega_i$,
and rewrite Eq.~\eqref{eq:symmetric_polynomial_1} as
\begin{align}
\label{eq:identity[10]}
2^{n-1}\sum_{(I, J); a_1=1}\Delta(I)^4\Delta(J)^4=\left[\sum_{(I, J); a_1 = 1}\Delta(I)^2\Delta(J)^2\right]^2,
\end{align}
where 
\begin{align}
\label{eq:Delta[1]}
&\Delta(\{a_1, a_2, \dots, a_r\})^l=\prod_{1\le p < q \le r}\left(x_{a_p}-x_{a_q}\right)^l,
\end{align}
and the summations are taken over 
\begin{align}
\label{eq:sum[1]}
&I=\{a_1, a_2\dots, a_n\}, ~1=a_1<a_2< \cdots <a_n,\\
\label{eq:sum[2]}
&J=\{b_{1}, b_2\dots, b_{n}\},~b_1<b_2<\cdots < b_n,\\
\label{eq:sum[3]}
&\{a_1, a_2, \dots, a_n, b_1, b_2, \dots, b_n\} \!=\! \{1, 2, \dots, 2n\}.
\end{align}
Equation~\eqref{eq:identity[10]} trivially holds for $n=1$. 
Below, we prove it for general $n$ by induction.

To this end, let us set 
\begin{align}
\label{eq:P[1]}
&P_n(x_1, x_2\dots, x_{2n}) 
\nonumber \\
&=\sum_{(I, J); a_1 =1 } \Delta(I)^4\Delta(J)^4\\
\label{eq:P[1-1]}
&=\sum_{(I', J')} \prod_{2\le p\le n}(x_1-x_{a_p})^4\Delta\left(I'\right)^4
\nonumber \\
&\times \prod_{2\le q\le n}(x_{b_1}-x_{b_q})^4\Delta\left(J'\right)^4,
\end{align}
and
\begin{align}
\label{eq:Q[1]}
&Q_n(x_1, x_2\dots, x_{2n}) 
\nonumber\\
&=\sum_{(I, J) ; a_1=1} \Delta(I)^2\Delta(J)^2\\
\label{eq:Q[1-1]}
&=\sum_{(I', J')}\prod_{2\le p\le n}(x_1-x_{a_p})^2\Delta\left(I'\right)^2 
\nonumber \\
&\times \prod_{2\le q\le n}(x_{b_1}-x_{b_q})^2\Delta\left(J'\right)^2,
\end{align}
where $I'=\{a_2, \dots, a_n\}$ and $J'=\{b_2, \dots, b_n\}$.
We assume the induction hypothesis, which is written in terms of $P_{n-1}$ and $Q_{n-1}$ as follows:
\begin{align}
\label{eq:induction_hypothesis}
2^{n-2}P_{n-1}= Q^2_{n-1}.
\end{align}

First, we prove that Eq.~(\ref{eq:induction_hypothesis}) holds for $x_1=x_2=x$.
Since $\Pi (x_1-x_{a_p})= 0 $ when $x_1=x_2=x$ and $a_2 =2$, we have 
\begin{align}
\label{eq:P[2-First]}
&P_n(x, x, x_3, \dots, x_{2n})\nonumber \\
&=\sum_{(I', J')} \prod_{2\le p\le n}(x-x_{a_p})^4\Delta\left(I'\right)^4 \prod_{2\le q\le n}(x-x_{b_q})^4\Delta\left(J'\right)^4\\
\label{eq:P[2-Second]}
&=\prod_{3\le k \le 2n}(x-x_k)^4\sum_{(I', J')}\Delta(I')^4\Delta(J')^4\\
\label{eq:P[2-Third]}
&=2\prod_{3\le k \le 2n}(x-x_k)^4 \sum_{(I', J'); a_2=3} \Delta(I')^4\Delta(J')^4 \\
\label{eq:P[2-Fourth]}
&=2\prod_{3\le k \le 2n}(x-x_k)^4 P_{n-1}(x_3,\dots, x_{2n})
\end{align} 
and
\begin{align}
\label{eq:Q[2]}
&Q_n(x, x, x_3, \dots, x_{2n})\nonumber \\
&= 2\prod _{3\le k \le 2n} (x-x_k)^2 Q_{n-1}(x_3, \dots, x_{2n}),
\end{align}
which, with help of the induction hypothesis in Eq.~\eqref{eq:induction_hypothesis}, 
give rise to  
\begin{align}
\label{eq:hypothesis[x1=x2=x]}
Q_{n}(x, x, x_3, \dots, x_{2n})^2 
=2^{n-1}P_{n}\left(x, x, x_3, \dots, x_{2n}\right).
\end{align}

This implies 
\begin{align}
\label{eq:hypothesis_n+1}
2^{n-1} P_n - Q_n^2  = (x_1 - x_2)^s {\cal R},
\end{align}
where the polynomial ${\cal R}$ is nonzero for $x_1=x_2=x$. 
In order to determine $s$, 
we prove 
\begin{align}
\label{eq:derivative_relation[1]}
2^{n-1}{\partial P_{n} (x, x, x_3\dots, x_{2n})\over \partial x}={\partial Q_{n} (x, x, x_3\dots, x_{2n})^2\over \partial x}. 
\end{align} 
By Eq.~\eqref{eq:P[2-Fourth]}, we have
\begin{align}
\label{eq:derivative_P[3]}
&{\partial P_{n} (x, x, x_3\dots, x_{2n})\over \partial x} \nonumber\\
&=\prod_{3\le l \le 2n}(x-x_l)^4 \sum_{3 \le k \le 2n}\!{8\over x-x_k}P_{n-1}(x_3, \dots, x_{2n}). 
\end{align}
Similarly, by Eq.~\eqref{eq:Q[2]}, we have 
\begin{align}
&{\partial Q_n (x, x, x_3, \dots, x_{2n})^2\over \partial x}\nonumber \\
\label{eq:derivative_Q[2_1]}
&=2Q_n(x, x, x_3, \dots, x_{2n})\\
\label{eq:derivative_Q[2_2]}
&\times\prod_{3\le l \le 2n}(x-x_l)^2\sum_{3 \le k \le 2n}{4\over x-x_k}Q_{n-1}(x_3, \dots, x_{2n})\\
\label{eq:derivative_Q[2_3]}
&=\prod_{2\le l \le 2n}(x-x_l)^4\sum_{3\le k \le 2n}{16\over x-x_k}Q_{n-1}(x_3, \dots, x_{2n})^2, 
\end{align}
which, with help of the induction hypothesis [Eq.~\eqref{eq:induction_hypothesis}] and Eq.~\eqref{eq:derivative_P[3]}, 
leads to the proof of Eq.~\eqref{eq:derivative_relation[1]}.
Thus, the partial derivative of both sides of Eq.~\eqref{eq:hypothesis_n+1} with respect to $x$ for $x_1=x_2=x$ 
gives rise to
\begin{align}
\label{eq:s[1]}
s(x_1-x_2)^{s-1}{\cal R}|_{x_1=x_2=x}=0,
\end{align}
which is satisfied only for $s\ge 2 $.
Hence, by the symmetry, we have
\begin{align}
\label{eq:s[2]}
2^{n-1}P_n - Q_n^2 = \prod_{1\le i < j \le 2n} (x_i - x_j)^2 {\cal V}
\end{align}
for some polynomial ${\cal V}$. 
However, the degree of each monomial in the left-hand side, which is $4n(n-1)$, 
is less than that in the right-hand side, which is at least $2n(2n-1)$. 
Therefore, ${\cal V}$ must be zero. 
Q.E.D.

%%%%%%%%%%%%%%%%%%%%%%%%%%%%%%%%%%%%%%%%%%%%%%%%%%%%%%%%%%%%%%%

%%%%%%%%%%%%%%%%%%%%%%%%%%%%%%%%%%%%%%%%%%%%%%%%%%%%%%%%%%%%%%%

\end{document}